\documentclass[%
reprint,
superscriptaddress,
 amsmath,amssymb,
 aps
]{revtex4-2}

\usepackage{graphicx}
\usepackage{dcolumn}
\usepackage{bm}
\usepackage{hyperref}

\begin{document}

\preprint{APS/123-QED}

\newcommand{\GMG}[1]{{\color{orange}{#1}}}
\newcommand{\cds}[1]{{\color{purple}{#1}}}
\newcommand{\mfh}[1]{{\color{red}{#1}}}
\newcommand{\bt}[1]{{\color{green}{#1}}}
\newcommand{\cmd}[1]{{\color{blue}{#1}}}
\newcommand{\dmh}[1]{{\color{yellow}{#1}}}

\newcommand*{\kT}{k_\text{B}T}
\newcommand*{\etaS}{\eta_{\rm s}}
\newcommand*{\etaB}{\eta_{\rm b}}
\newcommand*{\dB}{d_{\rm B}}
\newcommand*{\ns}{n_{\rm s}}

\preprint{APS/123-QED}

\title{Limits of economy and fidelity for programmable assembly of size-controlled triply-periodic polyhedra}

\author{Carlos M. Duque}
\affiliation{Max Planck Institute of Molecular Cell Biology and Genetics, 01307 Dresden, Germany}
\affiliation{Center for Systems Biology Dresden (CSBD), 01307 Dresden, Germany}
\affiliation{Department of Physics, University of Massachusetts, Amherst, MA 01003}
\author{Douglas M. Hall}
\affiliation{Department of Polymer Science and Engineering, University of Massachusetts, Amherst, MA 01003}
\author{Botond Tyukodi}
\affiliation{Department of Physics, Babes-Bolyai University, 400084 Cluj-Napoca, Romania}
\author{Michael F. Hagan}
\affiliation{Martin A. Fisher School of Physics, Brandeis University, Waltham MA 02453}
\author{Christian D. Santangelo}
\affiliation{Martin A. Fisher School of Physics, Brandeis University, Waltham MA 02453}
\author{Gregory M. Grason}
\affiliation{Department of Polymer Science and Engineering, University of Massachusetts, Amherst, MA 01003}

\date{\today}%

\begin{abstract}
We propose and investigate an extension of the Caspar-Klug symmetry principles for viral capsid assembly to the programmable assembly of size-controlled triply-periodic polyhedra, discrete variants of the Primitive, Diamond, and Gyroid cubic minimal surfaces.  Inspired by a recent class of programmable DNA origami colloids, we demonstrate that the economy of design in these crystalline assemblies -- in terms of the growth of the number of distinct particle species required with the increased size-scale (e.g. periodicity) -- is comparable to viral shells. We further test the role of geometric specificity in these assemblies via dynamical assembly simulations, which show that conditions for simultaneously efficient and high-fidelity assembly require an intermediate degree of flexibility of local angles and lengths in programmed assembly.  Off-target misassembly occurs via incorporation of a variant of disclination defects, generalized to the case of hyperbolic crystals.  The possibility of these topological defects is a direct consequence of the very same symmetry principles that underlie the economical design, exposing a basic tradeoff between design economy and fidelity of programmable, size controlled assembly.
\end{abstract}

\maketitle


\section{Introduction}.jpg
In Nature, self-assembly underlies the creation of functional materials. From photonic nanostructures~\cite{Thompson2004, Prum2009} to extracellular media~\cite{Neville1993, Fratzl2003a} to nanoencapsulation~\cite{Perlmutter2015, Kerfeld2018, Rother2016}, robust and dynamic control over the precise structure and size of these assemblies is essential to their adaptive properties. Living systems have evolved pathways to direct multi-protein assembly towards morphologies with a tunable size scale~\cite{Hagan2021}, from finite diameter shells and tubules~\cite{Oosawa1975} to periodically-modulated material composites~\cite{Bouligand2008}. Indeed, length scale is a fundamental but crucial element of structure control for functional materials. Yet, achieving structures with well-controlled lengths, particularly when the target size is much larger than the constituent size, poses a basic challenge. Notably, the ability to target triply-periodic (i.e. crystalline) architectures with specific symmetries and periodicities that can be scaled to larger than the subunit sizes is fundamental to control of desirable material functions, such as photonic bandgap behavior.  In particular, hybrid structures related to the Diamond and Gyroid surfaces are well-known to exhibit prominent bandgaps, and gyroid-like structures have evolved in diverse species of birds, beetles and butterflies as a means of producing structural coloration~\cite{Michielsen2008, Saranathan2010, Saranathan2021}. In these assemblies, targeted wavelength-selective properties are achieved by control over periodicity at the scale of 100s of {\rm nm}, at least an order of magnitude larger than the protein building blocks themselves.

\begin{figure*}[!t]
\centering\includegraphics[width=0.9\linewidth]{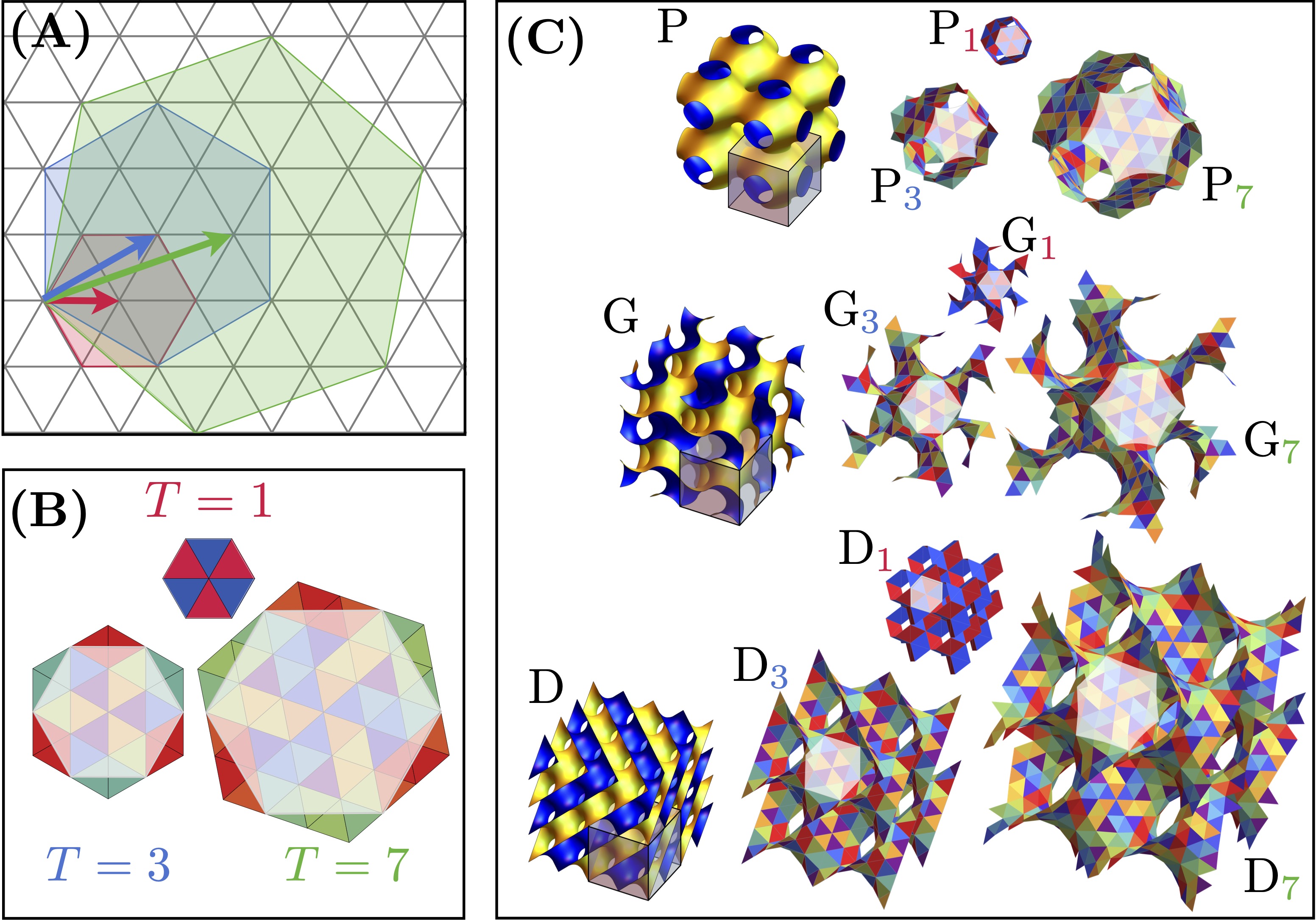}
\caption{\label{fig:fig1} {\bf Design of size-controlled triply-periodic minimal surface assembly from programmable triangular particles.}\textbf{(A)} A pair of integers $(h,k)$ is used to define a triangulation vector, ${\bf L}$, which connects a pair of vertices on a triangular lattice, defining the fundamental hexagonal patch. The hexagons in red, blue, and green are built with ${\bf L}$ vectors corresponding to the triangulation numbers $T=$ 1, 3 and 7. \textbf{(B)} Underlying triangular faces associated with each fundamental hexagon. Triangles with the same colors represent the same subunit type. \textbf{(C)} From top to bottom: $2\times2\times2$ subsets of primitive (P), gyroid (G), and diamond (D) TPMS. In each case, transparent boxes are used to highlight a single cubic unit-cell. Each cubic unit-cell is triangulated with the corresponding fundamental hexagon shown in panel \textbf{(B)}. The shaded areas on each triangulated unit cell highlight the fundamental hexagon used in the construction.}
\end{figure*}

These examples from nature have inspired recent efforts to design ``programmable'' building blocks to realize synthetic analogs of hierarchically-organized biological materials ~\cite{Ke2012, Jones2015, Zeravcic2017, Douglas2009, Huang2016}.  This strategy targets two key aspects of the protein building blocks of functional biological assemblies. First, new approaches to designing building block {\it geometry} from the molecular to the colloidal scale have enabled the fabrication of shapes of staggering complexity \cite{Glotzer2007, Sacanna2011, Heuckel2021, Su2020}. Second, encoding multiple species of subunits with specific interactions to favor a particular network of contacts allows the assignment of a specific  ``address'' to every assembled subunit \cite{Ke2012, Murugan2015, Murugan2015a, Jacobs2015, Jacobs2016, Zeravcic2017}. Yet, while such {\it addressable assembly} offers one simple and generic approach to this problem, where the number of interacting elements must grow with the target size, this paradigm suffers from a corresponding explosion in the complexity of multi-species mixtures as the target grows arbitrarily large\cite{Zeravcic2014, Ke2014, Ong2017}.  This is notably the case for programmable {\it crystalline} assemblies, where the unit cell dimensions typically scale with the size of the programmable building blocks themselves~\cite{Nykypanchuk2008, Auyeung2012, Jones2015, Tian2020, Rogers2011,Hensley2022,Majewski2021, Michelson2022}, so far limiting applications, for example, to the use of nanometric DNA based assembly units to target photonic structures with properties in the optical range~\cite{Park2020}.

DNA nanotechnology, in particular, has multiple strategies to implement  addressable assembly, and has been exploited to target and realize assemblies with a precisely defined, complex structures~\cite{Ke2014, Ong2017, Gerling2015}.
Recent works leverage the unprecedented combination of control over geometry and interaction specificity possible through DNA origami to realize a class of quasi-spherical shells and cylindrical tubes~\cite{Rothemund2004, Benson2015, Sigl2021, Videbaek2022, Hayakawa2022}.  Crucially, their target diameters are regulated by programming their curvature, achieving finite sizes that are much larger than the subunits.
The design strategy for size-controlled shells~\cite{Sigl2021} takes advantage of symmetry-based principles proposed for icosohedral viral shells~\cite{Zandi2004, Twarock2019, Siber2020, Johnson1997}, the celebrated {\it Caspar-Klug} (CK) construction~\cite{Caspar1962}.  The CK rules provide a rational means to determine the minimal number of inequivalent subunits (\textit{i.e.} conformations of capsid proteins) needed to form closed shells of arbitrarily large diameter, an economy of design presumably favored by selection pressures in viral evolution.  In this context, CK rules might be considered as one class of solutions to the generic problem of miminizing the {\it complexity} of specifically interacting subunit mixtures needed to achieve size-programmed assembly~\cite{Bohlin2023, Pinto2023}.

In this article we propose and explore the extension of the symmetry-based principles of CK to an entirely distinct class of programmable assemblies, triply-periodic polyhedra~\cite{Pedersen2018, Tanaka2023}, shown schematically in Fig. \ref{fig:fig1}.  Like tubules and shells, these are 2D surface-like assemblies of triangular subunits, in our case inspired by DNA origami assemblies~\cite{Sigl2021, Videbaek2022, Hayakawa2022}. In contrast to shell-like assemblies, our proposed design rules employ negative, rather than positive, Gaussian curvature.  In particular, we target the design of a related class of triply-periodic minimal surfaces (TPMS), the Gyroid (G), Diamond (D) and Primitive (P) surfaces~\cite{Lord2003, Schoen2012}, which all have cubic symmetry and negative Gaussian curvature.

We show here that the high symmetries of the P, G and D structures facilitate a similar economy of design as CK capsids, arising from the commensurability of their crystallographic space-groups with their decomposition into triangular building blocks.  Like CK designs of closed shells, we show how the combination of {\it interaction specificity}, encoded in specifically binding edge types, and {\it geometric specificity}, encoded by the target edge lengths of and dihedral angles between subunits, allows the programming of unit cell sizes that are, essentially, arbitrarily large compared to the subunit size.  We analyze the economy of design in triply-periodic polyhedral assemblies in terms of the scaling of the number of distinct subunits needed to achieve a given target periodicity in the crystal, and show that these ``inverted'' structures achieve similarly optimal scaling with increasing target size in comparison to icosohedral shells.

While the economy of design deriving from symmetry guarantees a unique target ground state, it does not guarantee that it correctly assembles, and if so, at reasonable rates or under physical conditions realizable in experiments.  Notably, recent experiments on self-closed tubular assemblies of DNA particles show that subunit flexibility, specifically dihedral bending, gives rise to significant off-target assembly into structures of undesired diameter \cite{Hayakawa2022}.  To understand how geometric specificity, in the form of angular and length flexibility of subunits, limits the ability to achieve controllable crystal dimensions through economically programmed assembly, we study grand canonical Monte Carlo simulations of a physical model of triangular subunit assembly. We consider the effects of variable elastic stiffness, and show that the range of subunit flexibility is restricted by the simultaneous requirements for high-fidelity and rapid assembly, which are respectively favored by low or high flexibility. Notably, the failure mechanism leading to misassembly for sufficiently flexible subunits, in the form of generalized disclination in hyperbolic crystalline structures, can be traced directly to the very same symmetry-based design that guarantees economical assembly.

\section*{Design economy of triply-periodic polyhedra}
\begin{figure}[!t]
\centering\includegraphics[width=1.0\linewidth]{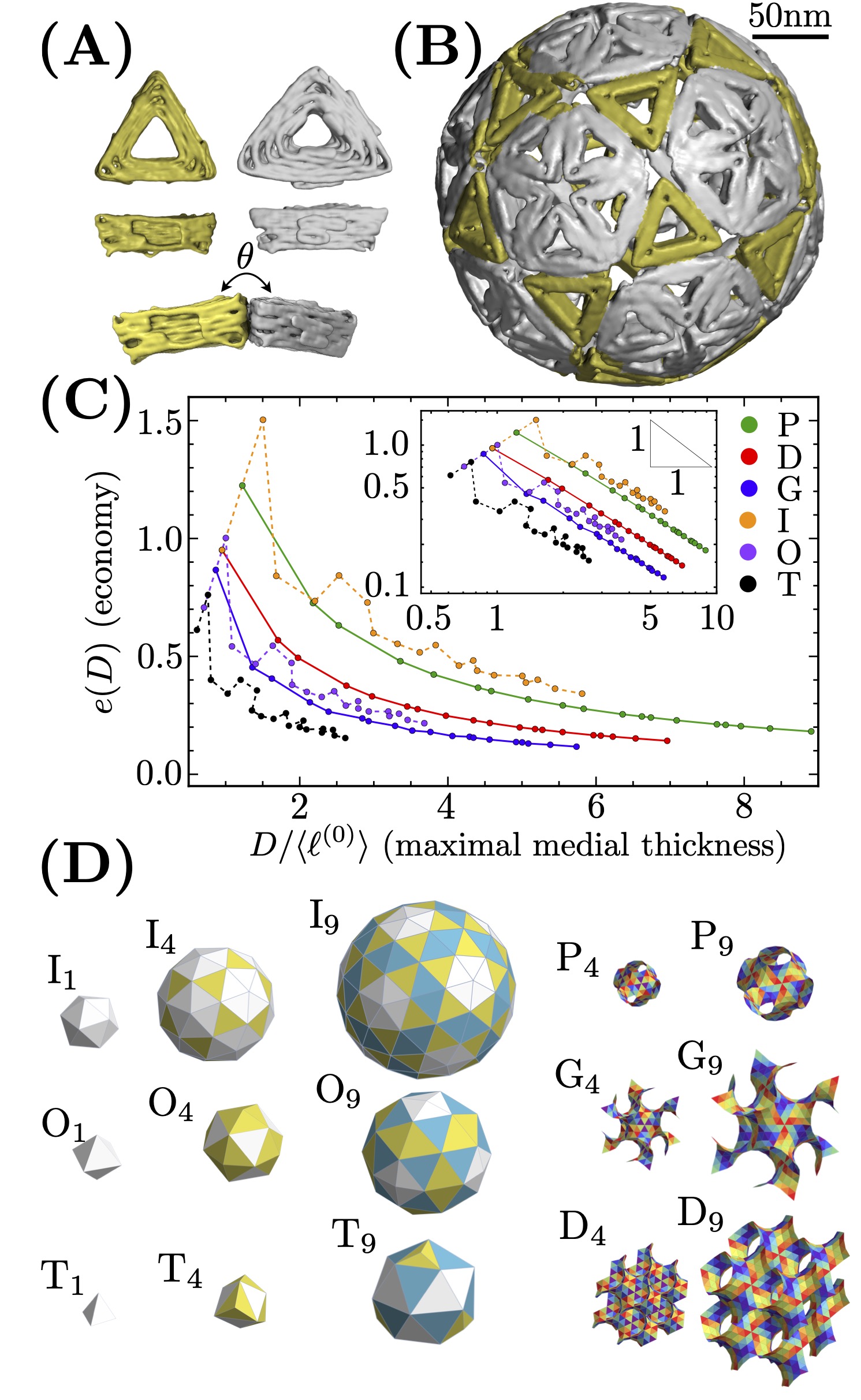}
\caption{\label{fig:fig2} {\bf Economy of programmable assembly of shells and TMPS via triangular particles}. \textbf{(A)} Cryo-eM reconstruction of the two different triangular DNA origami subunit types used in the self-assembly of $T=4$ icosahedral shell shown in \textbf{(B)} adapted from Sigl {\it et al.} ~\cite{Sigl2021}. The angle $\theta$ highlights the preferred dihedral angle between the subunits. \textbf{(C)} Measure of {\it economy}, $e(D)$ for polyhedral crystals (solid lines) and spherical shells (dashed-lines) as a function of the maximal medial thickness, $D$ (in units of mean edge length). Inset: Log-log plot of the same economy measures. \textbf{(D)} Triangulated shells of icosahedal, octahedral, and tetrahedral symmetries with triangulation number $T=$ 1, 4, and 9 built with $N_T=\lceil T/3\rceil$ distinct subunits (left).  $T=$  4 and 9 polyhedral crystals built with $N_T = T$ distinct subunits (right).}
\end{figure}

The economy of the CK construction for quasi-spherical assemblies stems from the fact that spherical shells can be decomposed into regular triangular ``subunits'', corresponding to geometric structures known as {\it deltahedra}.  In the original CK construction, subunits were themselves triplets of proteins that form the viral capsid~\cite{Caspar1962}.
Here, we follow the design of Sigl {\it et al} ~\cite{Sigl2021} and consider the triangular subunit as a single, self-assembling unit, shown schematically in Fig.~\ref{fig:fig2}A. The most economical shell designs are closed tilings of equilateral triangles, \textit{i.e.} the Platonic solids: tetrahedron (T), octahedron (O), and icosahedron (I).  Among these, the icosahedron possesses the most point symmetries, and correspondingly the largest number of equivalent triangular facets ($20$).  Structures composed of more than this number of triangles necessarily break the 3-fold symmetry of the equilateral subunit, and so increase the number of {\it inequivalent} triangular elements needed to form them.

CK argued that subtriangulations of the original triangular net that preserve icosahedral symmetry lead to the fewest symmetry-inequivalent positions on the closed shell, and thus require the fewest distinct subunits. Such subtriangulations are constructed from triangular subregions of a planar triangular lattice~\cite{Twarock2019, Siber2020}, and are parameterized by the lattice translation vector between vertices ${\bf L} = h{\bf a}_1 +k{\bf a}_2$, where $(h,k)$ are a pair of integers and ${\bf a}_1$,  ${\bf a}_2$ are basis vectors of the triangular lattice. Then $T\equiv |{\bf L}|^2 = h^2 + k^2 + h~k$ is the number of subtriangles per base triangle, resulting in a structure with $20 T$ subunits. However, the commensurability of subtriangulation and icoshedral symmetry implies that complete shells can be assembled from fewer distinct triangle types, $N_T$, which is equal to $\lceil T/3 \rceil$ for deltahedral shells (i.e. with equilateral base faces)~\footnote{There are $T$ inequivalent internal edges in a deltahedral tiling which are distributed into groupings of three (i.e. closed triangles).  The minimal number of distinct triplets is $\lceil T/3 \rceil$.  }. A design objective to maximize the target size, $D$, of an assembly for a minimal number of distinct subunit types, $N_T$, suggests a measure of {\it economy},
\begin{equation}
    e(D) \equiv D/N_T.
\end{equation}
As shown in Fig. \ref{fig:fig2}C, this measure decreases with target size as $e(D)\sim 1/D$ for deltahedral assemblies -- assemblies made from equilateral triangles -- where we use the maximal medial thickness~\cite{Turk2003} as the standard measure of size $D$.  This scaling can be understood from CK theory, since the triangulation number is proportional to the surface area and thus $T\propto D^2$. Notably, icosahedral shells maximize this measure of economy among deltahedral shells, which is also consistent with the CK logic.

\begin{figure}[!t]
\centering\includegraphics[width=1.0\linewidth]{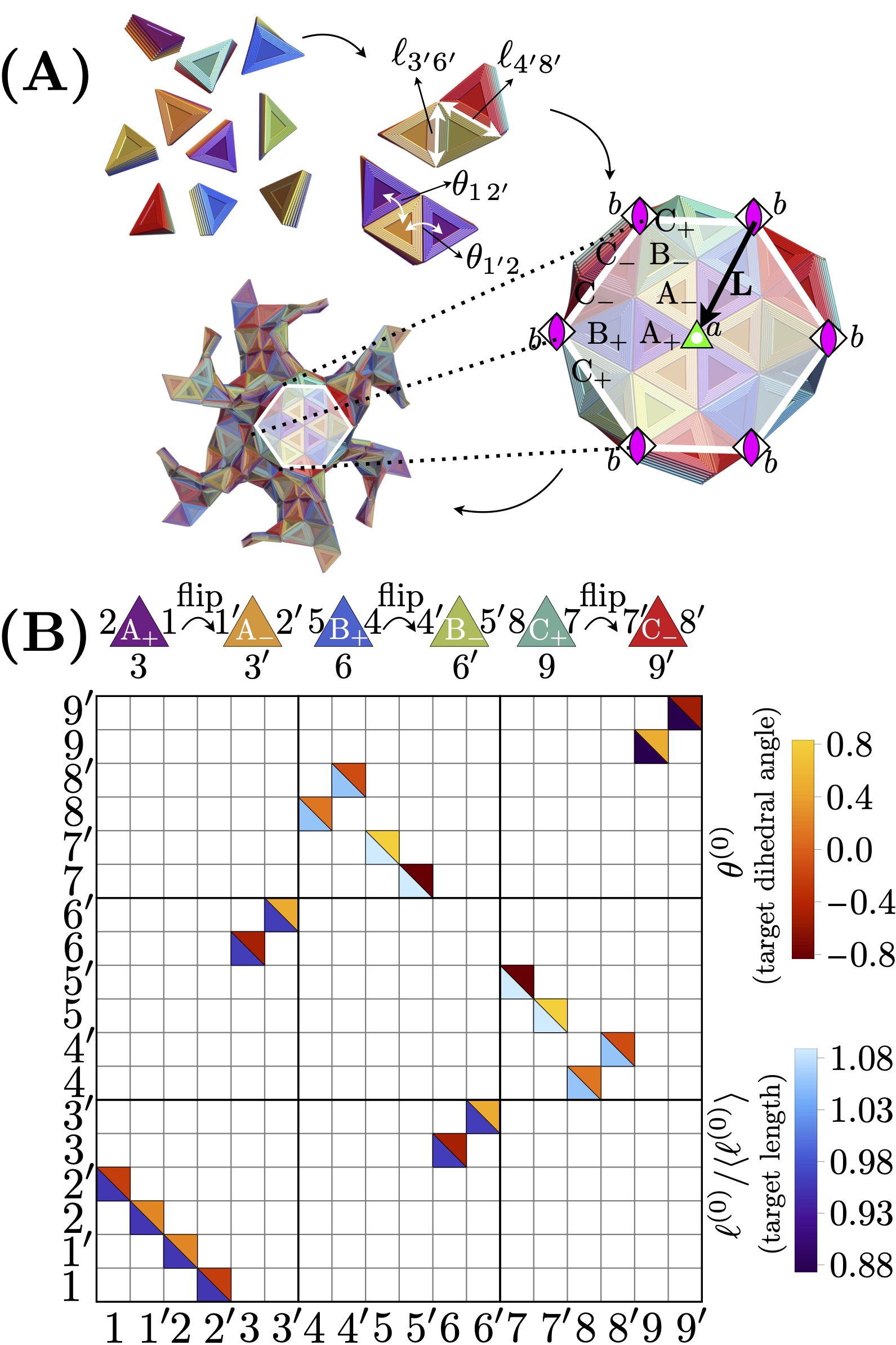}
\caption{\label{fig:fig3} {\bf Program for $T=3$ Gyroid assembly}.  \textbf{(A)} Illustration of the self-assembly process: triangular blocks of different types following their interaction matrix template form larger structures, such as the shown G$_3$ cubic unit-cell.  The three-dimensional rendering highlights the structure as DNA origami particle, with the important feature that opposite faces of the particle are distinct, as colored and referenced a $\pm$ in (B).  The center and vertices of the highlighted hexagon represent the Wyckoff sites present in the triangulation. The glyph next to each vertex denotes the symmetry of the Wyckoff site. Notice that the translation vector ${\bf L}$, which defines the $T$ number of the structure, joins Wyckoff sites of different symmetry. \textbf{(B)} Interaction matrix for G$_3$ with each colored block representing a valid edge-pairing. The monovalent design rules used in the construction ensure that a given edge-type is only allowed to bind to at most one other edge-type. The lower (upper) color of each colored-block of the matrix represents the target length (dihedral angle) of a valid edge-pairing. Both planar sides of each triangular block are colored differently to account for the ``flipping" symmetry exhibited by TPMS. This is accounted by assigning two edge-types to each triangular block planar side. Unprimed and primed edge-types are respectively the edge sides of the $+$ and $-$ planar sides of each triangular block.}
 \end{figure}

This design principle can be extended to triply-periodic triangular assemblies associated with the P, D, and G cubic minimal surfaces, by decomposing these structures into basic hexagonal elements, essentially following what has been dubbed ``hexagulation'' in studies of dense particle packings on these surfaces~\cite{Dotera2017}.  As developed by Sadoc and Charvolin~\cite{Sadoc1989} and elaborated by others~\cite{Ramsden2009, Pedersen2017, Pedersen2023}, high-symmetry tilings of P, G, and D can be derived from the $\{6,4\}$ tiling of the hyperbolic plane projected onto triply-periodic tesselations of $\mathbb{E}^3$ (Euclidean three space), provided that point symmetries at the center, vertices and edges of the hexagonal patches are preserved in the space group embedding.  However, while CK triangulations are constructed from triangular tilings meeting at 5-fold vertices, $\{6,4\}$ tilings are constructed from hexagonal tiles that meet in 4-fold vertices~\cite{Sadoc1989}, which is possible on the hyperbolic plane, as illustrated in Supporting Fig. S1~\footnote{Note that, strictly, the triangulation in the hyperplane satisfies rotoinversion symmetries on the vertices ($\bar{4}$) and centers ($\bar{3}$) of the fundamental hexagonal cells as annotated Supporting Fig. S1A.}.  When projected into cubic tesselations of $\mathbb{E}^3$, vertices of a projected skew hexagonal base ``tile'' are constrained to lie on specific Wyckoff positions of the corresponding crystallographic space group of the structure.  See for example, Fig. \ref{fig:fig3}A for G, where the central point of the shaded hexagon sits at the 16a position of $Ia\bar{3}d$, a point of 3-fold roto-inversion symmetry, while the six outer vertices sit at 24d, points of 4-fold roto-inversion~\cite{IntlTables}.  Like the triangular base elements of the CK construction, the hexagonal base tile of the triply-periodic P, G, and D polyhedra can be subtriangulated in a way that preserves the point symmetries of the $\{6,4\}$ tiling (see Methods and examples shown in Supporting Fig. S2). The corresponding triply-periodic triangulations are similarly triangulated by a vector ${\bf L}$ that connects a 4-coordinated vertex to center of the hexagon, and thus the tilings are  classified by  the $T$ number (Fig.~\ref{fig:fig1}A).   

We use these triply-periodic triangulations of $\mathbb{E}^3$ as the basis for triangular subunit designs that target the assembly of TPMS whose vertex positions map onto the corresponding minimal surface for arbitrary subtriangulation.  We denote these polyhedral target structures as P$_T$, D$_T$ and G$_T$ according to the respective cubic surface and $T$ number.  The geometric and topological data of the embedded structure --  edge length, connectivity and dihedral angles -- are then used to design target values for the subunit shape and interactions.  In our construction, the mean edge length of a basic triangular element is fixed for all structures, while the projection subtriangulation can introduce variation in edge lengths.  Fig. \ref{fig:fig3} shows an example for the G$_3$ structure in terms of 9 mutually interacting subunit edges. The interaction matrix in Fig. \ref{fig:fig3} includes both edge specificity and geometric data about the edge lengths and the dihedral angles formed when particular subunits meet. Notably, the coincident symmetries of the subtriangulation and the $Ia\bar{3}d$ space group allow for the large units cells ($96 T$ triangular particles per cubic repeat of G) to be constructed from only $N_T = T$ distinct subunits.  Supporting Figs. S3-6 show corresponding examples of binding specificity and geometry of edges for P, G and D for $T=1, 3$ and $7$

Unlike the CK structures, triply-periodic P$_T$, D$_T$ and G$_T$ polyhedra have infinite genus in their target (bulk) state. Nevertheless, each structure can be characterized by a well-defined, finite size, roughly corresponding to the characteristic pore size of the structure. To quantify and compare this programmable size scale, we computed the maximal medial thickness $D$ of each triply-periodic polyhedra~\cite{Turk2003}.  Fig. \ref{fig:fig2}C plots the finite size design economy, $e(D)$, as function of $T$ for  P$_T$, D$_T$ and G$_T$ polyhedra in comparison to generalized CK designs.  Like the CK assemblies, scaling the target size of the triply-periodic polyhedra to larger and larger dimensions (in units of the basic triangular elements) can be achieved at a similar level of economy, with $D \sim T^{1/2}$ and $N_T \sim T$, so that the power law scaling of $e$ with size is identical in the large size limit.


\begin{figure*}[!t]
\centering\includegraphics[width=1\linewidth]{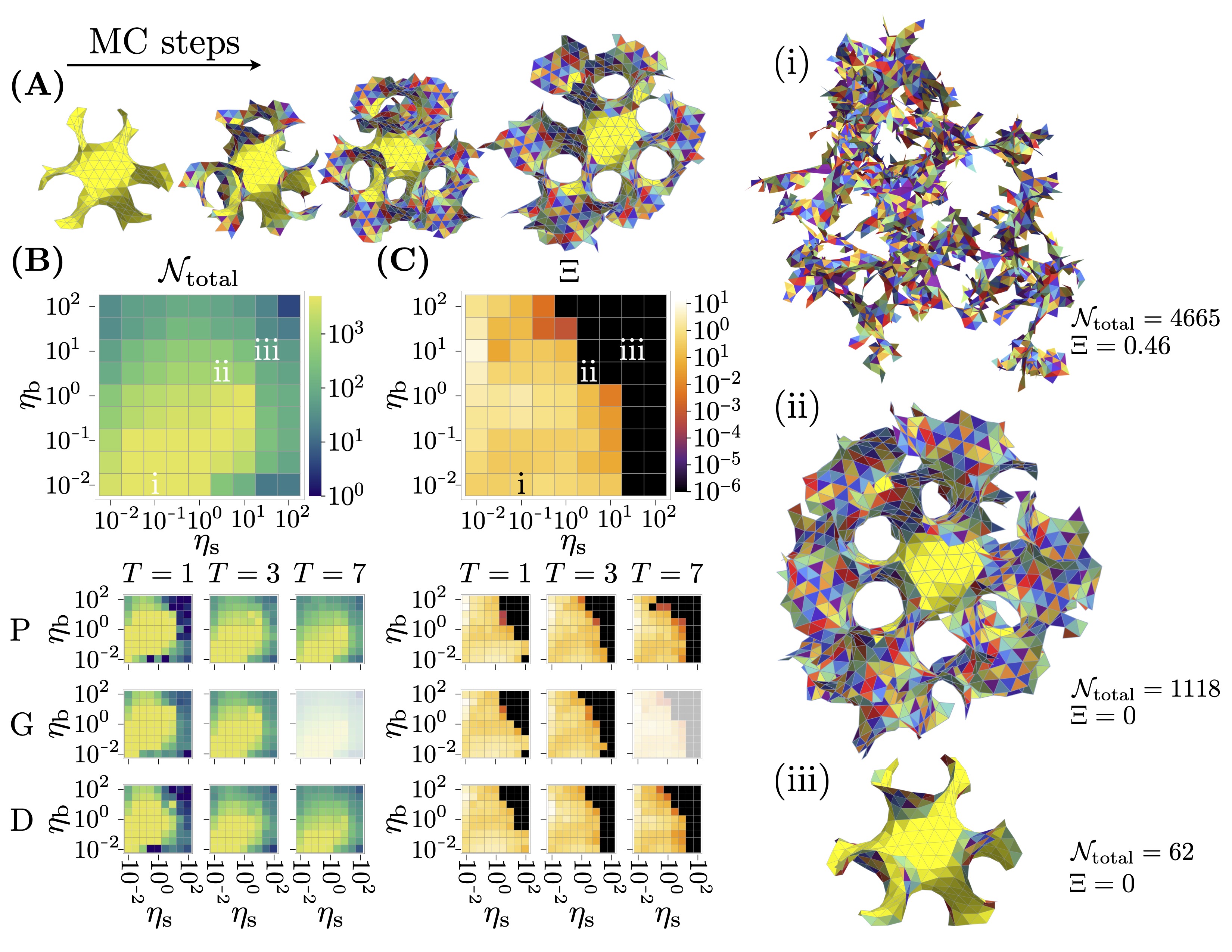}
\caption{\label{fig:fig4} {\bf Efficiency and fidelity of assembly versus geometric specificity of binding.} \textbf{(A)} An example assembly simulated trajectory of the $G_7$ assembly, growing from a fixed ``seed'' (yellow), with newly bond subunits shown in corresponding colors.  (B) and (C), respectively, show the number of assembled subunits (${\cal N}_{\text{total}}$) and residual shape strain ($\Xi$) from simulation trajectories for a range of dimensionless bending ($\eta_{\bf b}$) and stretching ($\eta_{\bf s}$) stiffness, the upper panel highlights the results for $G_7$, while results for all three symmetries and $T=1,3$ and $7$ are shown below.  Points (i), (ii) and (iii) highlight three different conditions for $G_7$ assembly corresponding the final structures shown on the right:  (i) rapid, off-target assembly; (ii) productive, on-target assembly; and (iii) non-productive, rigid assembly (also shown in Supporting Movies S4-6).
}
\end{figure*}

\section*{Efficiency versus fidelity of size-economic assembly}
The economy of design of our construction for triply-periodic polyhedra depends on the combination of both the interaction specificity between different edge types and the geometric specificity of the edge-edge contacts. To understand the physical limits for the fidelity and assembly yield of these structures, we implemented a coarse-grained model of triangular particle assembly employed in Tyukodi {\it et al.}~\cite{Tyukodi2022}, in which assembled structures are triangular meshes, with degrees of freedom at their vertices.  Similar elastic mesh simulations have been applied to model assemblies of shells and tubules~\cite{Rotskoff2018,Li2018,Panahandeh2020,Tyukodi2022,Fang2022,Mohajerani2022}.  Here, the energy of an assembly derives from interactions of bound edges indexed by $i$ and $j$,
\begin{equation}
\label{eq: energy}
    E= \sum_{\langle i j\rangle}\Big\{ - \epsilon_{ij} + \frac{k}{2}| \ell_{ij} - \ell^{(0)}_{ij}|^2 + \kappa \big[ 1 - \cos (\theta_{ij} - \theta_{ij}^{(0)}) \big] \Big\} .
\end{equation}
The first term describes the binding energy of the edges. We assume that edge $i$ and edge $j$ have a common binding affinity, $\epsilon_{ij}= \epsilon_{\rm bind}$ if they are programmed to interact, but do not bind otherwise. Notably, this interaction specificity, which is crucial for forming a target triply-periodic polyhedra, is an important contrast to previous studies of capsid assembly in which all subunit interactions were identical \cite{Rotskoff2018,Wagner2015a,Li2018,Li2019,Panahandeh2020}, but is analogous to models of capsid assembly in which subunit interactions followed CK rules \cite{Schwartz1998,Hagan2006,Jack2007,Hagan2011, Rapaport2008,Rapaport2010,Rapaport2012,Whitelam2009,Nguyen2007,Nguyen2008,Nguyen2009,Wilber2007,Wilber2009,Cheng2012, Mohajerani2022}.
The second and third terms in eq. (\ref{eq: energy}) describe the energy cost of edge stretching and bending, respectively, where $\ell^{(0)}_{ij}$ and $\theta_{ij}^{(0)}$ are the target values for edge $ij$, as determined from the geometrical embedding of P$_T$, D$_T$ and G$_T$ (e.g. as shown for G$_3$ in Fig. \ref{fig:fig3}).  The respective moduli for edge stretching and dihedral bending between bound faces are given by $k$ and $\kappa$. To consider the influence of these distinct elastic modes on assembly behavior, we introduce the dimensionless ratios relative to binding,
\begin{equation}
    \etaS \equiv k \langle |\ell^{(0)}_{ij}|^2 \rangle/\epsilon_{\rm bind}; \ \ \etaB \equiv \kappa/\epsilon_{\rm bind}
\end{equation}
where $ \langle |\ell^{(0)}_{ij}|^2 \rangle$ is the mean-square target edge length.  

To model near-equilibrium assembly behavior, we perform grand canonical Monte Carlo simulations (see Methods), which consider a single cluster of bound, triangular units held at fixed chemical potential with respect to a population of free subunits (i.e. monomers) composed of a mixture of all the triangular species needed to assemble a given triply-periodic polyhedron.  To test the efficiency and quality of targeted assembly, an initial seed of the preassembled structure is prepared, and the MC algorithm considers three types of moves:  1) addition/removal of free subunits to an appropriately unbound triangle edge of the cluster; 2) vertex displacement of assembled particles; and 3) fission/fusion of edges between bound/unbound edges of the assembled cluster.  Moves are accepted with the Boltzmann-weighted probability according to the change in energy, Eq. (\ref{eq: energy}), and chemical potential $\mu$ for free monomer addition at fixed temperature.  Example simulation trajectories are shown for $T=3$ structures in Supporting Movies S1-3.  

To consider the role of geometric specificity, we choose $\epsilon_{\rm bind} = -6.5 \kT$ and chemical potential $\mu = -4.5 \kT$ and vary the elastic/binding ratios. We introduce ${\cal N}_{\rm total}$, the average number of assembled units in the particle cluster, as a measure of assembly efficiency. To capture the fidelity of the assembly we quantify the mean-quadratic strain 
\begin{equation}
\Xi = \langle | \ell_{ij} - \ell^{(0)}_{ij}|^2 \rangle/\langle |\ell^{(0)}_{ij}|^2 \rangle+ \langle|\theta_{ij} - \theta_{ij}^{(0)}|^2\rangle,
\end{equation}
which is computed from the elastic ground-state of the ultimate structure to remove the influence of thermal fluctuations.  We terminated simulations at $50\times10^6$ MC sweeps, or when ${\cal N}_{\rm total} =5000$.  We simulated assembly trajectories for P, G, and D structures for a range of $T$-numbers and varying the elastic/binding ratios over four orders of magnitude: $\eta_{\text{s}}\in [10^{-2},10^{2}]$ and $\eta_{\text{b}} \in [10^{-2},10^{2}]$.

Fig. \ref{fig:fig4}B and C show simulation results for the dependence of ${\cal N}_{\rm total}$ and $\Xi$ on $\etaB$ and $\etaS$ for G$_7$ assemblies.  They show that ${\cal N}_{\rm total}$ decreases with both stretching and bending stiffness, with the efficiency near zero outside of the region $\etaS\lesssim 10$ and $\etaB\lesssim 1$.  In contrast, the high-flexibility  regime is the regime of {\it low fidelity} as indicated by the large, non-zero residual strain values $\Xi$.  These $\Xi > 0$ values are evidence of assemblies that form with the correct edge matching specificity, but nevertheless have a topology that is incompatible with the target edge lengths and dihedral angles.  
In other words, while assembly is rapid when subunits are flexible, $\etaS \ll 1$ or $\etaB \ll 1$ result in highly defective, ``off-target'' structures, as shown for a structure like Fig. \ref{fig:fig4}i.  In the opposite limit of $\etaS \gg 1$ or $\etaB \gg 1$, while bonds form the correct geometry so that $\Xi \simeq 0$, the assembly efficiency is low due to the small rate of new subunits joining to a free edge. This results in a structure like Fig. \ref{fig:fig4}iii, with few if any additional subunits bound to the seed.  However, at intermediate flexibility -- approximately $5\lesssim \etaS\lesssim 10 $ and $5\lesssim \etaB\lesssim 10 $ -- assembly achieves both significant yield (i.e. ${\cal N}_{\rm total}\gtrsim 10^2$) and high fidelity (i.e. $\Xi\simeq 0$), indicating productive and defect-free assembly of the target crystalline structure (Fig. \ref{fig:fig4}ii).

The smaller panels of Fig. \ref{fig:fig4}B and C  compare the assembly efficiency, ${\cal N}_{\rm total}$, and fidelity, $\Xi$, for all three P$_T$, G$_T$, and D$_T$ structures and for sequences of increasing target sizes, corresponding to $T=1, 3$, and $7$.  All cases show the same qualitative dependence on angular and length flexibility of bound subunits: rapid yet off-target assembly at high flexibility, on-target yet sluggish assembly for stiff structures, and a regime of productive, on-target assembly in the intermediate flexibility regime.  These overall trends reveal that interaction specificity alone is not sufficient for reaching target assemblies. This is indeed consistent with the fact that P$_T$, G$_T$ and D$_T$ have an identical interaction matrix for each given $T$, but differ in terms of the target edge lengths and dihedral angles.

\begin{figure*}[!t]
\centering\includegraphics[width=1.0\linewidth]{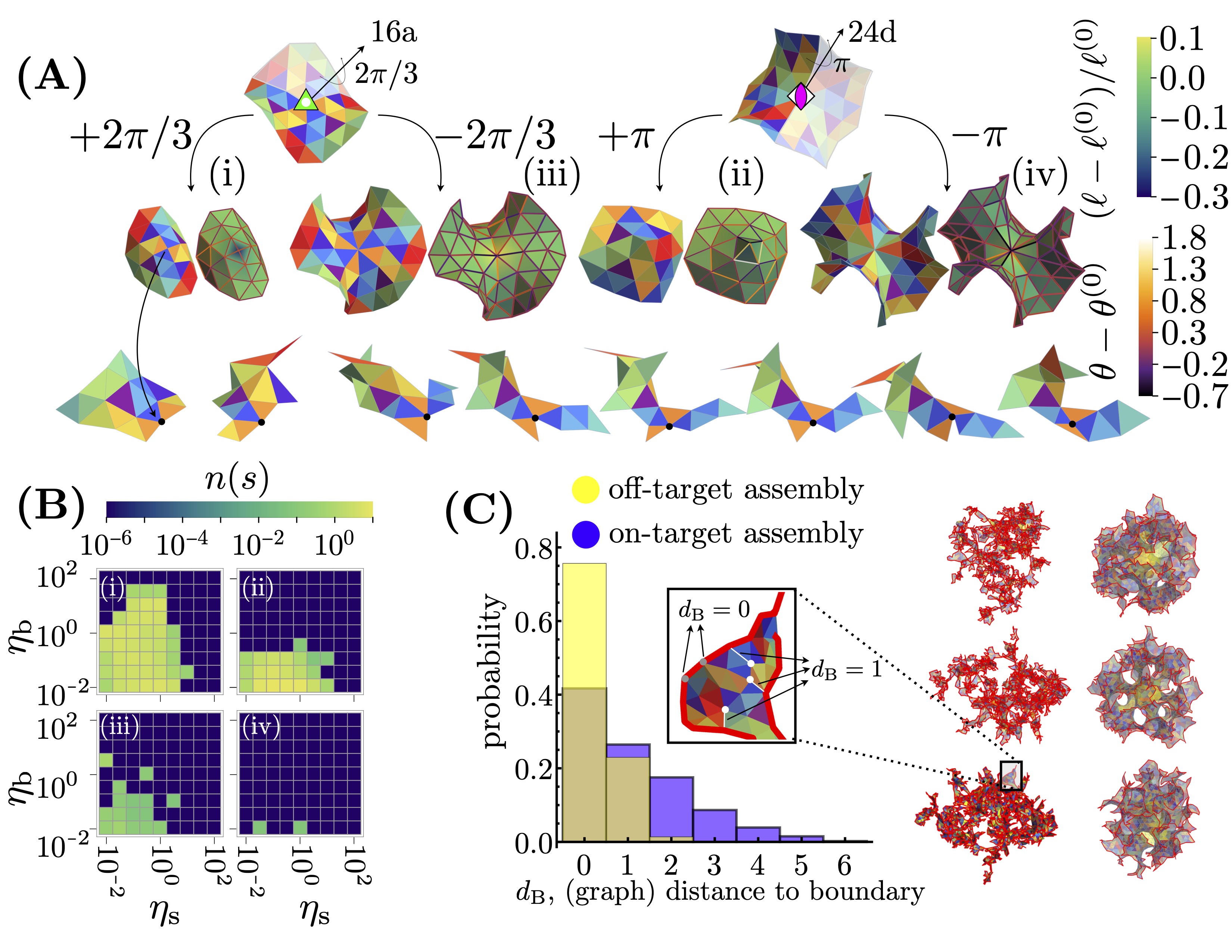}
\caption{\label{fig:fig5} {\bf Disclination pathways to off-target assembly.} \textbf{(A)} Positive and negative topological defects compatible with the self-assembly matching rules of a G$_7$ structure, corresponding respectively to fractional edge removed or added from the target assembly, extending from a high-symmetry vertex. Here, possible defects are constructed by a Volterra-like construction on defectless patches centered at the Wyckoff sites 16a and 24d, where the topological charge $s$ quantifies the excess/defecitive of rotation angle around the discliation. Each defective patch is colored with respect to the unique triangle species as shown on the left patches as well as their length strain (faces) and angle strain (edges) as shown on the right patches. The snapshot sequence on the bottom shows a possible assembly pathway of a topological defect with charge $s=+2\pi/3$. \textbf{(B)} Number of disclinations formed per primitive cell, $n(s)$, as a function of dimensionless ratios $\etaS$ and $\etaB$ for each of the defect types labeled from (i) to (iv) for simulated $G_7$ assembly. \textbf{(C)} Blue (Yellow): probability of a vertex to have a graph distance, $\dB$, to the boundary of a structure exhibiting on-target (off-target) assembly. The structure snapshots on the right (left) column show different views of the on-target (off-target) structures. The enlarged box shows different examples of vertices having a graph distance $\dB=0$ and $\dB=1$. Off- and in-target assembly correspond to points (i) and (ii) in Fig.~\ref{fig:fig4}, respectively.}
\end{figure*}

It is instructive to compare these observations to results of models for  positive curvature capsids. While we find that interaction specificity (edge-type binding specificity) is essential for assembly of target triply-periodic polyhedral structures, small capsid structures can assemble without interaction specificity, i.e. from systems of identical subunits, within certain ranges of bending and stretching moduli \cite{Rotskoff2018,Wagner2015a,Li2018,Panahandeh2020}. However, to form larger  ($T>7$) capsids with icosahedral symmetry, interaction specificity \cite{Sigl2021} or templating \cite{Li2018,Sigl2021} is essential, and interaction specificity significantly increases target yields and robustness to parameter variations for smaller capsids \cite{Mohajerani2022}.  Our observation that, in addition to interaction specificity, a minimal level of geometric specificity is essential to form a target triply-periodic polyhedral structure is also consistent with capsids. Even when interaction specificity allows for only a single ground state capsid structure, malformed structures assemble under conditions of low geometric specificity and/or strong interactions because mis-bound subunits do not have time to anneal before becoming trapped in the assembly \cite{Schwartz1998,Hagan2006,Jack2007,Hagan2011, Rapaport2008,Rapaport2010,Rapaport2012,Whitelam2009,Nguyen2007,Nguyen2008,Nguyen2009,Wilber2007,Wilber2009,Cheng2012,Hagan2014,Whitelam2015,Mohajerani2022}. On the other hand, too much geometric specificity leads to low kinetic cross-sections and thus slow assembly \cite{Whitelam2009,Hagan2006}.

\section*{Defect-mediated mis-assembly} In our  model  of triply-periodic polyhedra, bound subunits have perfect type-specificity, and hence, off-target assemblies must have the same local network of subunits but the wrong global geometry. Here, we show that the primary mechanism of mis-assembly derives from topological defects of the quasi-2D crystalline assembly.  These defects take the form of point disclinations, defined relative to the target polyhedral assembly (see Fig. \ref{fig:fig5}A), with an angular wedge of the triangular mesh removed or added relative to its ideal geometry as one encircles a vertex.  As in the standard convention for wedge disclinations in 2D crystals~\cite{Seung1988}, we associate the topological charge $s$ of a disclination to the excess degree of bond rotation around a vertex relative to the target structure. This charge can be defined and measured at any given vertex (see Methods and Supporting Fig. S7).  As edge binding only takes place between complementary edge-types, such disclinations are only possible at vertices of rotational symmetry in the target assembly, i.e. at vertices located at Wyckoff positions in the target assembly.  This means, for example for the $Ia \bar{3}d$ space group of the G$_7$ structure, the $\bar{3}$ symmetry of the 16a position supports $s=\pm 2 \pi/3$ disclinations, while the $\bar{4}$ symmetry of the 24d position supports $s=\pm \pi$ disclinations (\ref{fig:fig5}A).

Fig. \ref{fig:fig5}B shows the average number of defects per unit cell $n(s)$ for different disclination types in simulations of $T=7$ assemblies, for the same range of flexibilities considered in Fig. \ref{fig:fig4}. Notably, these defects appear  when bonds are flexible, coincident with the regime of large residual strains $\Xi$ in off-target assembly, indicating that disclination formation is the primary mechanism of mis-assembly. Surprisingly,  this defect population is biased in the {\it sign} of topological charge.  That is, defects with either $s=+ 2 \pi/3$ or $+\pi$ charge form at finite density, corresponding to patches with wedges {\it removed} relative to the stress-free target geometry, but the negatively-charged variants of these defects do not form at significant densities even in highly bendable or stretchable assemblies for these parameters~\footnote{Our definition of disclination charge accounts for the effective negative Gaussian curvature of the target crystal, but defines excess bond rotation relative to a target bond coordination that may be larger than 6 (e.g. the 8-coordinated vertex at position 24d of G structures).}.  This same bias is also observe for P, G and D assemblies for variable $T$ numbers, as shown in Supporting Fig. S8.   Similarities between the local elastic energies of positive and negative disclination types in Supporting Fig. S9A suggest that this imbalance is not driven by differences in strains generated by these defects types. Instead, we find by simulation of variable excluded volume sizes (Supporting Fig. S9B) that the bias towards $s=+ 2 \pi/3$ or $+\pi$ relative to their negative counterparts is a feature of steric interactions between triangular units:  volume exclusion tends to penalize crowding {\it excess} triangular units around a shared vertex required by $s<0$ defects.


Finally, we note that the residual net topological charge of defects has consequences for the gross morphology of off-target assemblies.  In a 2D crystal that grows isotropically, a finite disclination charge density would tend to generate elastic energies that grow superextensively, i.e. faster than the number of subunits~\cite{Seung1988}.  Instead, we find a morphological transition in defective assemblies that mitigates growth of elastic energy with assembly. Fig. \ref{fig:fig5}C compares the graph distance $\dB$ of vertices to a {\it free boundary} in the assembly for the conditions corresponding to off-target vs. on-target assembly in Fig.\ref{fig:fig4}, points (i) and (ii) respectively. Notably, off-target mis-assembly results in narrow, stringy structures ($\dB \lesssim 2$), while on-target assembly of sufficiently rigid subunits results in bulk 2D assembly with the free boundary extending far away from interior subunits. The stringy morphology of off-target mis-assembly has the important effect of reducing the far-field elastic cost of disclinations, as the long-range effects of defects are screened by the presence of free boundaries~\cite{Grason2012, Li2019}.  Hence, analogous to the anisotropic domain growth in curvature-frustrated 2D crystals~\cite{Schneider2005, Meng2014, Grason2016} or filaments~\cite{Yang2010}, we argue that this narrow, strip-like morphology eludes the superextensive costs that would be otherwise be generated by finite disclination charge densities in isotropically growing 2D crystals.

\section*{Discussion}

In summary, we have extended the economical design principles of the CK construction of closed shells to triply-periodic, negative curvature programmable assembly.  In both cases, economy derives from the commensurabilty of the symmetries of a subtriangulation with the symmetry elements of the target structure.  For triply-periodic polyhedra, this requires constraining the vertices and centers of the base tile to the Wyckoff sites of appropriate symmetry.  Preserving these symmetries in the sub-triangulation guarantees that the subtriangulation is composed of ``redundant'' copies of relatively few symmetry-inequivalent particles.  

However, while the high-symmetry is necessary for design economy, it is also  the source of off-target misassembly.  Notably, the very same rotational symmetries that anchor the sub-tilings of P$_T$, D$_T$ and G$_T$ are sites where disclinations are possible, and these disclinations proliferate if the geometric specificity of the binding between subunits is too low.  Indeed, as the example in Supporting Fig. S10 illustrates, certain $T$ values ($T= 4,\,12,\,16,\ldots$) lead to an additional set of (2-fold) Wyckoff positions, enabling the formation of a third set of $n \pi$ disclinations and thus more assembly errors.  This trade-off between design economy and the propensity for misassembly is unavoidable. It leads to narrowly defined regimes where assembly is flexible enough to occur at reasonable rates, but specific enough to suppress disclinations.  The design criteria for efficient and high-fidelity assembly (i.e. $5\lesssim \etaS\lesssim 10 $ and $5\lesssim \etaB\lesssim 10 $) are thus critical for the experimental design and realization of size-controlled crystals.  In the context of programmable DNA triangles, experimental yields of off-target tubule assembly suggest a range of bending stiffness to binding ratio in the range $\etaB \approx 0.1-1$, and dimensional arguments suggest $\etaS$ to be in a similar range, which is notably in the range of productive and high-fidelity assembly of our physical model \cite{Hayakawa2022}.

These estimates suggest this system may be ideal for harnessing the economy of P$_T$, D$_T$ and G$_T$ for programmed assembly of crystal structures with unit cells that are tunable to dimensions larger than those of the subunits, which is currently a challenge with assembly of colloidal or supramolecular building blocks.  This limitation applies to current approaches to program the assembly of complex crystals of DNA functionalized particles~\cite{Rogers2011,Hensley2022}, as well as DNA origami ``voxels''~\cite{Majewski2021, Michelson2022}.  Notably, the nanometric size of programmable building blocks typically puts the photonic bandgap behavior far outside of the range of the optical regime for DNA-programmable crystals, as unit cells sizes have been typically limited to within a few times the subunit size.  Hence it would be advantageous to use P$_T$, D$_T$ and G$_T$ assemblies as platforms for bottom-up design of photonic materials, with wavelength tunable via $T$.  For example, taking computed bandgaps for gyroidal crystals~\cite{Saranathan2011} and using the $\sim 50$ nm size of DNA origami triangles~\cite{Sigl2021}, suggests that photonic behavior occurs in the visible range for $T= 4 - 9$.

Finally, we note that recent approaches to synthetic protein design have assembled icosohedral shells of highly-modular size and structure, realizing CK structures in the range of $T=4 - 100$~\cite{Dowling2023}.  Thus, we anticipate that our ``inverted CK'' design principles could be  a template for engineering new classes of triply-periodic, protein-based mesoporous frameworks of controllable periodicity and symmetry.

\section{Acknowledgements}
The authors thank S. Fraden and D. Hayakawa for help with rendering of cryoEM reconstructions DNA origami assemblies and also B. Rogers for many useful comments on this manuscript.  This work was primarily support by the NSF through the Brandeis Center for Bioinspired Soft Materials, an NSF MRSEC (DMR-2011846).  The authors acknowledge additional support from the German Federal Ministry of Education and Research under grant number 031L0160 and European Union's Horizon 2020 Research and Innovation Programme under grant agreement no. 829010 (CMD); European Union’s Horizon 2020 research and innovation programme under the Marie Skłodowska-Curie grant agreement No 101026118 (BT); NSF grant DMR-2309635 (MFH); and NSF grant DMR-2217543 (CDS). Simulations where performed using the UMass Cluster at the Massachusetts Green High Performance Computing Center and the High Performance Computing Cluster of MPI-CBG and CSBD.  The authors dedicate this article to memory and legacy of A. Schoen.

\section{Supplementary Information}

\noindent {\bf Supplementary Information} - Supplementary Figures and Appendices.

\noindent \hypertarget{V1}{{\bf Supplementary Video 1}} (\url{http://www.pse.umass.edu/sites/default/files/grason/images/p3.mov})) - Self-assembly trajectory for a P$_3$ structure in the on-target regime with $\etaS=10$, $\etaB=5$, $\mu=-4.5 k_B T$, and $\epsilon_{\text{bind}}=-6.5 k_B T$ The yellow subunits form a minimal translational simple cubic unit-cell which we use as the seed for self-assembly.

\noindent \hypertarget{V2}{{\bf Supplementary Video 2}} ( \url{http://www.pse.umass.edu/sites/default/files/grason/images/g3.mov})- Self-assembly trajectory for a G$_3$ structure in the on-target regime with $\etaS=10$, $\etaB=5$, $\mu=-4.5 k_B T$, and $\epsilon_{\text{bind}}=-6.5 k_B T$ The yellow subunits form a minimal translational bcc unit-cell which we use as the seed for self-assembly.

\noindent \hypertarget{V3}{{\bf Supplementary Video 3}}  (\url{http://www.pse.umass.edu/sites/default/files/grason/images/d3.mov}) - Self-assembly trajectory for a D$_3$ structure in the on-target regime with $\etaS=10$, $\etaB=5$, $\mu=-4.5 k_B T$, and $\epsilon_{\text{bind}}=-6.5 k_B T$ The yellow subunits form a minimal translational fcc unit-cell which we use as the seed for self-assembly.

\noindent \hypertarget{V4}{{\bf Supplementary Video 4}}   (\url{http://www.pse.umass.edu/sites/default/files/grason/images/g7\_i.mov}) - Self-assembly trajectory for a G$_7$ structure in the off-target regime corresponding to the structure shown on Fig. 4i with $\etaS=0.1$, $\etaB=0.01$, $\mu=-4.5 k_B T$, and $\epsilon_{\text{bind}}=-6.5 k_B T$ The yellow subunits form a minimal translational bcc unit-cell which we use as the seed for self-assembly.

\noindent \hypertarget{V5}{{\bf Supplementary Video 5}}  (\url{http://www.pse.umass.edu/sites/default/files/grason/images/g7\_ii.mov}) - Self-assembly trajectory for a G$_7$ structure in the on-target regime corresponding to the structure shown on Fig. 4ii with $\etaS=5$, $\etaB=5$, $\mu=-4.5 k_B T$, and $\epsilon_{\text{bind}}=-6.5 k_B T$ The yellow subunits form a minimal translational bcc unit-cell which we use as the seed for self-assembly.

\noindent \hypertarget{V6}{{\bf Supplementary Video 6}}  (\url{http://www.pse.umass.edu/sites/default/files/grason/images/g7\_iii.mov}) - Self-assembly trajectory for a G$_7$ structure in the unefficient regime corresponding to the structure shown on Fig. 4iii with $\etaS=50$, $\etaB=10$, $\mu=-4.5 k_B T$, and $\epsilon_{\text{bind}}=-6.5 k_B T$ The yellow subunits form a minimal translational bcc unit-cell which we use as the seed for self-assembly.

\appendix

\section{Triply-periodic triangulations and interaction rules}  
Our construction of triangulations of $P$, $G$ and $D$ are based on projecting portions of planar triangular graphs on the level-set models of these minimal surfaces, in a way the preserves the symmetries of the *246 tiling of the $\mathbb{H}^2$ in the respective cubic space groups of $\mathbb{E}^3$~\cite{Ramsden2009}.  In brief, this construction begins with a triangular base, 1/6 of the hexagonal patch.  The vertices this triangular base, denoted ${\bf a}, {\bf b}$ and ${\bf c}$, are constrained to lie on Wyckoff site positions with appropriate rotoinversion symmetries to embedded the *246 tiling.  As shown in Table \ref{tab:wyckoff-table}, ${\bf a}$ is placed at the $\bar{3}$ center of hexagonal patch, while ${\bf b}$ and ${\bf c}$ lie on $\bar{4}$ points, corresponding to vertex where four hexagonal patches meet.  Last, we note that inversion ``flips'' the normal to triangular particles, so that this hexagonal base is therefore constructed by a single symmetry equivalent unit.  This base triangle itself, when embedded into the respective $Im\bar{3}m$, $Ia\bar{3}d$ and $Pn\bar{3}m$ space groups thus constitutes the $T=1$ triangulation of P, G and D.

Higher $T$ number triangulations follow from a procedure (described in detail in SI Text Sec. 1) where by the planar base triangle ${ {\bf a}, {\bf b}, {\bf c}}$, is subtriangulated according the identical construction as CK, followed by a projection of the vertex positions from the planar bases (arranged according to the space group symmetries) onto a level set model of P, G or D via simple gradient flow.

This procedure, in general, results in distortions of the dihedral angles between edge-sharing triangular faces, as well as lengths of edges, geometric information which we then record and define to set the {\it target values} on triangular subunits and their selective interactions (see SI Text Sec. 2).

\begin{table}
\centering
\begin{tabular}{l|l|l} \toprule
    {\text{TPMS}} & {${\bf a}$} & {${\bf b},\,{\bf c}$} \\ 
    P ($Im\bar{3}m$)  & $8c$ & $12d$ \\
    G ($Ia\bar{3}d$)  & $16a$ & $24d$ \\
    D ($Pn\bar{3}m$)  & $4c$ & $6d$ \\ 
\end{tabular}
\caption{\label{tab:wyckoff-table}Wyckoff site symmetries and locations for the vertices the base triangulation tile (i.e. 1/6 of the fundamental hexagon).  Notice that vertices ${\bf b}$ and ${\bf c}$ share the same Wyckoff site symmetry.}
\end{table}


\section{Grand Canonical MC simulation}
For the assembly simulations, we use a model previously developed for icosahedral shell self assembly \cite{Rotskoff2018,Zandi2004,Zandi2006,Zandi2020} and then adapted by us for arbitrary triangle design \cite{Tyukodi2022,Videbaek2022}. Subunits in the model are flexible triangles which can bind to each other along an edge. The local preferred curvature is modelled by preferred dihedral angles between neighboring faces sharing a bond (edge) and any deviation from the preferred dihedral angle has a corresponding bending energy cost. The energy  associated to triangle-triangle binding, edge stretching and dihedral change (bending) are shown in Eq. \ref{eq: energy}.

Each triangular subunit consists of 3 edges, each of which may be of distinct interaction type. The interaction matrix defines which type is allowed to bind to which type. If there are $\ns$ subunit species in the simulation, there are at most $3 \ns$ edge types with $(3 \ns)^2$ different interactions, i.e. binding energies, bending moduli and dihedral angles. In addition, each of the $3 \ns$ edge types may have their own stretching moduli and rest lengths. In the simulations presented in the main text, we fixed the binding energies for all allowed edge pair types to the same value. Similarly, bending moduli for all pairs and stretching moduli for all edge types are also set to the same value. Moreover, each allowed edge type pair has its own dihedral angle.

The simulation follows the growth of a single structure in the grand canonical ensemble, i.e. the structure is immersed in a bath of freely diffusing subunits held at fixed concentration. Concentrations (or, equivalently, the chemical potentials) for all species are set to be equal. The dynamics is governed by a series of Monte Carlo moves with fixed relative rates. The moves allow for subunit exchange between the structure and the bath, internal binding and unbinding of edges and thermal fluctuation of vertices with no change in topology. There are a total number of 11 moves and each move is carefully designed to satisfy detailed balance with its reverse move \cite{Tyukodi2022}.

\section{Calculation of defect charges}

Given a structure assembled with the matching rules of triply-periodic polyhedra, we can determine disclination charges by considering closed paths around encircling the disclinations like the one shown on the Supporting Fig. S7. Each path can be in general seen as a series of steps in which each individual step consists of a composition of two rotations: an initial rotation of angle $\phi_{ij}$ corresponding to the angle between two consecutive edges $E_i$ and $E_j$ and a second rotation of angle $\theta_{j}$ corresponding to the dihedral angle associated with the edge $E_j$ (s). This approach closely follows the formalism introduced by belcastro and Hull in which origami folding patterns are viewed as collections of affine transformations around the internal vertices of the patterns \cite{Belcastro2002}. Furthermore, moving around the vertex or rotating the surface around the vertex are analogous operations so we can perform all the $\phi$ and $\theta$ rotations around axes passing through the enclosed vertex and parallel to $\hat{\mathbf{z}}$ and $\hat{\mathbf{x}}$ respectively. One full rotation amounts then to the composition of rotation operations (from right to left) $\hat{R}_{v}(\Phi,\hat{\mathbf{n}})$:
\begin{eqnarray}
\hat{R}_v(\Phi,\hat{\mathbf{n}})&=&\hat{R}\left(\theta_0,\,\hat{\mathbf{x}}\right)\hat{R}\left(\phi_{n0},\,\hat{\mathbf{z}}\right)\ldots\nonumber\\
&&\hat{R}\left(\theta_2,\,\hat{\mathbf{x}}\right)\hat{R}\left(\phi_{12},\,\hat{\mathbf{z}}\right)\hat{R}\left(\theta_1,\,\hat{\mathbf{x}}\right)\hat{R}\left(\phi_{01},\,\hat{\mathbf{z}}\right),
\end{eqnarray}
where $\Phi$ is the desired angle around disclination vertex $v$ whose absolute value can be found can be determined as $\left|\Phi\right|=\arccos[(\text{Tr}(\hat{R}_{v})-1)/2]$. With the rotation angle $\Phi$ the disclination charge of a defect is defined as the angle deficit  $s=2\pi-\left|\Phi\right|$.

\bibliography{pnas-sample,all-references-self-assembly.2023.05.09}

\begin{thebibliography}{99}%
\makeatletter
\providecommand \@ifxundefined [1]{%
 \@ifx{#1\undefined}
}%
\providecommand \@ifnum [1]{%
 \ifnum #1\expandafter \@firstoftwo
 \else \expandafter \@secondoftwo
 \fi
}%
\providecommand \@ifx [1]{%
 \ifx #1\expandafter \@firstoftwo
 \else \expandafter \@secondoftwo
 \fi
}%
\providecommand \natexlab [1]{#1}%
\providecommand \enquote  [1]{``#1''}%
\providecommand \bibnamefont  [1]{#1}%
\providecommand \bibfnamefont [1]{#1}%
\providecommand \citenamefont [1]{#1}%
\providecommand \href@noop [0]{\@secondoftwo}%
\providecommand \href [0]{\begingroup \@sanitize@url \@href}%
\providecommand \@href[1]{\@@startlink{#1}\@@href}%
\providecommand \@@href[1]{\endgroup#1\@@endlink}%
\providecommand \@sanitize@url [0]{\catcode `\\12\catcode `\$12\catcode
  `\&12\catcode `\#12\catcode `\^12\catcode `\_12\catcode `\%12\relax}%
\providecommand \@@startlink[1]{}%
\providecommand \@@endlink[0]{}%
\providecommand \url  [0]{\begingroup\@sanitize@url \@url }%
\providecommand \@url [1]{\endgroup\@href {#1}{\urlprefix }}%
\providecommand \urlprefix  [0]{URL }%
\providecommand \Eprint [0]{\href }%
\providecommand \doibase [0]{https://doi.org/}%
\providecommand \selectlanguage [0]{\@gobble}%
\providecommand \bibinfo  [0]{\@secondoftwo}%
\providecommand \bibfield  [0]{\@secondoftwo}%
\providecommand \translation [1]{[#1]}%
\providecommand \BibitemOpen [0]{}%
\providecommand \bibitemStop [0]{}%
\providecommand \bibitemNoStop [0]{.\EOS\space}%
\providecommand \EOS [0]{\spacefactor3000\relax}%
\providecommand \BibitemShut  [1]{\csname bibitem#1\endcsname}%
\let\auto@bib@innerbib\@empty
\bibitem [{\citenamefont {Thompson}\ and\ \citenamefont
  {Parker}(2004)}]{Thompson2004}%
  \BibitemOpen
  \bibfield  {author} {\bibinfo {author} {\bibfnamefont {J.~M.~T.}\
  \bibnamefont {Thompson}}\ and\ \bibinfo {author} {\bibfnamefont {A.~R.}\
  \bibnamefont {Parker}},\ }\bibfield  {title} {\bibinfo {title} {A vision for
  natural photonics},\ }\href {https://doi.org/10.1098/rsta.2004.1458}
  {\bibfield  {journal} {\bibinfo  {journal} {Philosophical Transactions of the
  Royal Society of London. Series A: Mathematical, Physical and Engineering
  Sciences}\ }\textbf {\bibinfo {volume} {362}},\ \bibinfo {pages} {2709}
  (\bibinfo {year} {2004})},\ \Eprint
  {https://arxiv.org/abs/https://royalsocietypublishing.org/doi/pdf/10.1098/rsta.2004.1458}
  {https://royalsocietypublishing.org/doi/pdf/10.1098/rsta.2004.1458}
  \BibitemShut {NoStop}%
\bibitem [{\citenamefont {Prum}\ \emph {et~al.}(2009)\citenamefont {Prum},
  \citenamefont {Dufresne}, \citenamefont {Quinn},\ and\ \citenamefont
  {Waters}}]{Prum2009}%
  \BibitemOpen
  \bibfield  {author} {\bibinfo {author} {\bibfnamefont {R.~O.}\ \bibnamefont
  {Prum}}, \bibinfo {author} {\bibfnamefont {E.~R.}\ \bibnamefont {Dufresne}},
  \bibinfo {author} {\bibfnamefont {T.}~\bibnamefont {Quinn}},\ and\ \bibinfo
  {author} {\bibfnamefont {K.}~\bibnamefont {Waters}},\ }\bibfield  {title}
  {\bibinfo {title} {{Development of colour-producing $\beta$-keratin
  nanostructures in avian feather barbs}},\ }\bibfield  {journal} {\bibinfo
  {journal} {Journal of The Royal Society Interface}\ }\textbf {\bibinfo
  {volume} {6}},\ \href {https://doi.org/10.1098/rsif.2008.0466.focus}
  {10.1098/rsif.2008.0466.focus} (\bibinfo {year} {2009})\BibitemShut {NoStop}%
\bibitem [{\citenamefont {Neville}(1993)}]{Neville1993}%
  \BibitemOpen
  \bibfield  {author} {\bibinfo {author} {\bibfnamefont {A.~C.}\ \bibnamefont
  {Neville}},\ }\href {https://doi.org/10.1017/CBO9780511601101} {\emph
  {\bibinfo {title} {{Biology of Fibrous Composites}}}}\ (\bibinfo  {publisher}
  {Cambridge University Press},\ \bibinfo {address} {Cambridge},\ \bibinfo
  {year} {1993})\BibitemShut {NoStop}%
\bibitem [{\citenamefont {Fratzl}(2003)}]{Fratzl2003a}%
  \BibitemOpen
  \bibfield  {author} {\bibinfo {author} {\bibfnamefont {P.}~\bibnamefont
  {Fratzl}},\ }\bibfield  {title} {\bibinfo {title} {{Cellulose and collagen:
  from fibres to tissues}},\ }\href
  {https://doi.org/10.1016/S1359-0294(03)00011-6} {\bibfield  {journal}
  {\bibinfo  {journal} {Current Opinion in Colloid {\&} Interface Science}\
  }\textbf {\bibinfo {volume} {8}},\ \bibinfo {pages} {32} (\bibinfo {year}
  {2003})}\BibitemShut {NoStop}%
\bibitem [{\citenamefont {Perlmutter}\ and\ \citenamefont
  {Hagan}(2015)}]{Perlmutter2015}%
  \BibitemOpen
  \bibfield  {author} {\bibinfo {author} {\bibfnamefont {J.~D.}\ \bibnamefont
  {Perlmutter}}\ and\ \bibinfo {author} {\bibfnamefont {M.~F.}\ \bibnamefont
  {Hagan}},\ }\bibfield  {title} {\bibinfo {title} {Mechanisms of virus
  assembly},\ }\href {https://doi.org/10.1146/annurev-physchem-040214-121637}
  {\bibfield  {journal} {\bibinfo  {journal} {Annual Review of Physical
  Chemistry}\ }\textbf {\bibinfo {volume} {66}},\ \bibinfo {pages} {217}
  (\bibinfo {year} {2015})},\ \bibinfo {note} {pMID: 25532951},\ \Eprint
  {https://arxiv.org/abs/https://doi.org/10.1146/annurev-physchem-040214-121637}
  {https://doi.org/10.1146/annurev-physchem-040214-121637} \BibitemShut
  {NoStop}%
\bibitem [{\citenamefont {Kerfeld}\ \emph {et~al.}(2018)\citenamefont
  {Kerfeld}, \citenamefont {Aussignargues}, \citenamefont {Zarzycki},
  \citenamefont {Cai},\ and\ \citenamefont {Sutter}}]{Kerfeld2018}%
  \BibitemOpen
  \bibfield  {author} {\bibinfo {author} {\bibfnamefont {C.~A.}\ \bibnamefont
  {Kerfeld}}, \bibinfo {author} {\bibfnamefont {C.}~\bibnamefont
  {Aussignargues}}, \bibinfo {author} {\bibfnamefont {J.}~\bibnamefont
  {Zarzycki}}, \bibinfo {author} {\bibfnamefont {F.}~\bibnamefont {Cai}},\ and\
  \bibinfo {author} {\bibfnamefont {M.}~\bibnamefont {Sutter}},\ }\bibfield
  {title} {\bibinfo {title} {Bacterial microcompartments},\ }\href
  {https://doi.org/10.1038/nrmicro.2018.10} {\bibfield  {journal} {\bibinfo
  {journal} {Nature Reviews Microbiology}\ }\textbf {\bibinfo {volume} {16}},\
  \bibinfo {pages} {277} (\bibinfo {year} {2018})}\BibitemShut {NoStop}%
\bibitem [{\citenamefont {Rother}\ \emph {et~al.}(2016)\citenamefont {Rother},
  \citenamefont {Nussbaumer}, \citenamefont {Renggli},\ and\ \citenamefont
  {Bruns}}]{Rother2016}%
  \BibitemOpen
  \bibfield  {author} {\bibinfo {author} {\bibfnamefont {M.}~\bibnamefont
  {Rother}}, \bibinfo {author} {\bibfnamefont {M.~G.}\ \bibnamefont
  {Nussbaumer}}, \bibinfo {author} {\bibfnamefont {K.}~\bibnamefont
  {Renggli}},\ and\ \bibinfo {author} {\bibfnamefont {N.}~\bibnamefont
  {Bruns}},\ }\bibfield  {title} {\bibinfo {title} {{Protein Cages and
  Synthetic Polymers: A Fruitful Symbiosis for Drug Delivery Applications,
  Bionanotechnology and Materials science}},\ }\href
  {https://doi.org/10.1039/C6CS00177G} {\bibfield  {journal} {\bibinfo
  {journal} {Chem. Soc. Rev.}\ }\textbf {\bibinfo {volume} {45}},\ \bibinfo
  {pages} {6213} (\bibinfo {year} {2016})}\BibitemShut {NoStop}%
\bibitem [{\citenamefont {Hagan}\ and\ \citenamefont
  {Grason}(2021)}]{Hagan2021}%
  \BibitemOpen
  \bibfield  {author} {\bibinfo {author} {\bibfnamefont {M.~F.}\ \bibnamefont
  {Hagan}}\ and\ \bibinfo {author} {\bibfnamefont {G.~M.}\ \bibnamefont
  {Grason}},\ }\bibfield  {title} {\bibinfo {title} {Equilibrium mechanisms of
  self-limiting assembly},\ }\href
  {https://doi.org/10.1103/RevModPhys.93.025008} {\bibfield  {journal}
  {\bibinfo  {journal} {Reviews of Modern Physics}\ }\textbf {\bibinfo {volume}
  {93}},\ \bibinfo {pages} {025008} (\bibinfo {year} {2021})},\ \bibinfo {note}
  {publisher: American Physical Society}\BibitemShut {NoStop}%
\bibitem [{\citenamefont {Oosawa}\ and\ \citenamefont
  {Asakura}(1975)}]{Oosawa1975}%
  \BibitemOpen
  \bibfield  {author} {\bibinfo {author} {\bibfnamefont {F.}~\bibnamefont
  {Oosawa}}\ and\ \bibinfo {author} {\bibfnamefont {S.}~\bibnamefont
  {Asakura}},\ }\href@noop {} {\emph {\bibinfo {title} {{Thermodynamics of the
  Polymerization of Protein}}}}\ (\bibinfo  {publisher} {Academic Press},\
  \bibinfo {address} {London},\ \bibinfo {year} {1975})\BibitemShut {NoStop}%
\bibitem [{\citenamefont {Bouligand}(2008)}]{Bouligand2008}%
  \BibitemOpen
  \bibfield  {author} {\bibinfo {author} {\bibfnamefont {Y.}~\bibnamefont
  {Bouligand}},\ }\bibfield  {title} {\bibinfo {title} {Liquid crystals and
  biological morphogenesis: Ancient and new questions},\ }\href
  {https://doi.org/https://doi.org/10.1016/j.crci.2007.10.001} {\bibfield
  {journal} {\bibinfo  {journal} {Comptes Rendus Chimie}\ }\textbf {\bibinfo
  {volume} {11}},\ \bibinfo {pages} {281} (\bibinfo {year} {2008})}\BibitemShut
  {NoStop}%
\bibitem [{\citenamefont {Michielsen}\ and\ \citenamefont
  {Stavenga}(2008)}]{Michielsen2008}%
  \BibitemOpen
  \bibfield  {author} {\bibinfo {author} {\bibfnamefont {K.}~\bibnamefont
  {Michielsen}}\ and\ \bibinfo {author} {\bibfnamefont {D.}~\bibnamefont
  {Stavenga}},\ }\bibfield  {title} {\bibinfo {title} {Gyroid cuticular
  structures in butterfly wing scales: biological photonic crystals},\ }\href
  {https://doi.org/10.1098/rsif.2007.1065} {\bibfield  {journal} {\bibinfo
  {journal} {Journal of The Royal Society Interface}\ }\textbf {\bibinfo
  {volume} {5}},\ \bibinfo {pages} {85} (\bibinfo {year} {2008})},\ \Eprint
  {https://arxiv.org/abs/https://royalsocietypublishing.org/doi/pdf/10.1098/rsif.2007.1065}
  {https://royalsocietypublishing.org/doi/pdf/10.1098/rsif.2007.1065}
  \BibitemShut {NoStop}%
\bibitem [{\citenamefont {Saranathan}\ \emph {et~al.}(2010)\citenamefont
  {Saranathan}, \citenamefont {Osuji}, \citenamefont {Mochrie}, \citenamefont
  {Noh}, \citenamefont {Narayanan}, \citenamefont {Sandy}, \citenamefont
  {Dufresne},\ and\ \citenamefont {Prum}}]{Saranathan2010}%
  \BibitemOpen
  \bibfield  {author} {\bibinfo {author} {\bibfnamefont {V.}~\bibnamefont
  {Saranathan}}, \bibinfo {author} {\bibfnamefont {C.~O.}\ \bibnamefont
  {Osuji}}, \bibinfo {author} {\bibfnamefont {S.~G.~J.}\ \bibnamefont
  {Mochrie}}, \bibinfo {author} {\bibfnamefont {H.}~\bibnamefont {Noh}},
  \bibinfo {author} {\bibfnamefont {S.}~\bibnamefont {Narayanan}}, \bibinfo
  {author} {\bibfnamefont {A.}~\bibnamefont {Sandy}}, \bibinfo {author}
  {\bibfnamefont {E.~R.}\ \bibnamefont {Dufresne}},\ and\ \bibinfo {author}
  {\bibfnamefont {R.~O.}\ \bibnamefont {Prum}},\ }\bibfield  {title} {\bibinfo
  {title} {{Structure, function, and self-assembly of single network gyroid
  (I4132) photonic crystals in butterfly wing scales}},\ }\href
  {https://doi.org/10.1073/pnas.0909616107} {\bibfield  {journal} {\bibinfo
  {journal} {Proceedings of the National Academy of Sciences}\ }\textbf
  {\bibinfo {volume} {107}},\ \bibinfo {pages} {11676} (\bibinfo {year}
  {2010})}\BibitemShut {NoStop}%
\bibitem [{\citenamefont {Saranathan}\ \emph
  {et~al.}(2021{\natexlab{a}})\citenamefont {Saranathan}, \citenamefont
  {Narayanan}, \citenamefont {Sandy}, \citenamefont {Dufresne},\ and\
  \citenamefont {Prum}}]{Saranathan2021}%
  \BibitemOpen
  \bibfield  {author} {\bibinfo {author} {\bibfnamefont {V.}~\bibnamefont
  {Saranathan}}, \bibinfo {author} {\bibfnamefont {S.}~\bibnamefont
  {Narayanan}}, \bibinfo {author} {\bibfnamefont {A.}~\bibnamefont {Sandy}},
  \bibinfo {author} {\bibfnamefont {E.~R.}\ \bibnamefont {Dufresne}},\ and\
  \bibinfo {author} {\bibfnamefont {R.~O.}\ \bibnamefont {Prum}},\ }\bibfield
  {title} {\bibinfo {title} {{Evolution of single gyroid photonic crystals in
  bird feathers}},\ }\href {https://doi.org/10.1073/pnas.2101357118} {\bibfield
   {journal} {\bibinfo  {journal} {Proceedings of the National Academy of
  Sciences}\ }\textbf {\bibinfo {volume} {118}},\ \bibinfo {pages}
  {e2101357118} (\bibinfo {year} {2021}{\natexlab{a}})}\BibitemShut {NoStop}%
\bibitem [{\citenamefont {Ke}\ \emph {et~al.}(2012)\citenamefont {Ke},
  \citenamefont {Ong}, \citenamefont {Shih},\ and\ \citenamefont
  {Yin}}]{Ke2012}%
  \BibitemOpen
  \bibfield  {author} {\bibinfo {author} {\bibfnamefont {Y.}~\bibnamefont
  {Ke}}, \bibinfo {author} {\bibfnamefont {L.~L.}\ \bibnamefont {Ong}},
  \bibinfo {author} {\bibfnamefont {W.~M.}\ \bibnamefont {Shih}},\ and\
  \bibinfo {author} {\bibfnamefont {P.}~\bibnamefont {Yin}},\ }\bibfield
  {title} {\bibinfo {title} {{Three-Dimensional Structures Self-Assembled from
  DNA Bricks}},\ }\href {https://doi.org/10.1126/science.1227268} {\bibfield
  {journal} {\bibinfo  {journal} {Science}\ }\textbf {\bibinfo {volume}
  {338}},\ \bibinfo {pages} {1177} (\bibinfo {year} {2012})}\BibitemShut
  {NoStop}%
\bibitem [{\citenamefont {Jones}\ \emph {et~al.}(2015)\citenamefont {Jones},
  \citenamefont {Seeman},\ and\ \citenamefont {Mirkin}}]{Jones2015}%
  \BibitemOpen
  \bibfield  {author} {\bibinfo {author} {\bibfnamefont {M.~R.}\ \bibnamefont
  {Jones}}, \bibinfo {author} {\bibfnamefont {N.~C.}\ \bibnamefont {Seeman}},\
  and\ \bibinfo {author} {\bibfnamefont {C.~A.}\ \bibnamefont {Mirkin}},\
  }\bibfield  {title} {\bibinfo {title} {{Programmable materials and the nature
  of the DNA bond}},\ }\href {https://doi.org/10.1126/science.1260901}
  {\bibfield  {journal} {\bibinfo  {journal} {Science}\ }\textbf {\bibinfo
  {volume} {347}},\ \bibinfo {pages} {1260901} (\bibinfo {year}
  {2015})}\BibitemShut {NoStop}%
\bibitem [{\citenamefont {Zeravcic}\ \emph {et~al.}(2017)\citenamefont
  {Zeravcic}, \citenamefont {Manoharan},\ and\ \citenamefont
  {Brenner}}]{Zeravcic2017}%
  \BibitemOpen
  \bibfield  {author} {\bibinfo {author} {\bibfnamefont {Z.}~\bibnamefont
  {Zeravcic}}, \bibinfo {author} {\bibfnamefont {V.~N.}\ \bibnamefont
  {Manoharan}},\ and\ \bibinfo {author} {\bibfnamefont {M.~P.}\ \bibnamefont
  {Brenner}},\ }\bibfield  {title} {\bibinfo {title} {{Colloquium : Toward
  living matter with colloidal particles}},\ }\href
  {https://doi.org/10.1103/RevModPhys.89.031001} {\bibfield  {journal}
  {\bibinfo  {journal} {Reviews of Modern Physics}\ }\textbf {\bibinfo {volume}
  {89}},\ \bibinfo {pages} {031001} (\bibinfo {year} {2017})}\BibitemShut
  {NoStop}%
\bibitem [{\citenamefont {Douglas}\ \emph {et~al.}(2009)\citenamefont
  {Douglas}, \citenamefont {Dietz}, \citenamefont {Liedl}, \citenamefont
  {H\"ogberg}, \citenamefont {Graf},\ and\ \citenamefont {Shih}}]{Douglas2009}%
  \BibitemOpen
  \bibfield  {author} {\bibinfo {author} {\bibfnamefont {S.~M.}\ \bibnamefont
  {Douglas}}, \bibinfo {author} {\bibfnamefont {H.}~\bibnamefont {Dietz}},
  \bibinfo {author} {\bibfnamefont {T.}~\bibnamefont {Liedl}}, \bibinfo
  {author} {\bibfnamefont {B.}~\bibnamefont {H\"ogberg}}, \bibinfo {author}
  {\bibfnamefont {T.}~\bibnamefont {Graf}},\ and\ \bibinfo {author}
  {\bibfnamefont {W.}~\bibnamefont {Shih}},\ }\bibfield  {title} {\bibinfo
  {title} {Self-assembly of dna into nanoscale three-dimensional shapes},\
  }\href {https://doi.org/10.1038/nature08016} {\bibfield  {journal} {\bibinfo
  {journal} {Nature}\ }\textbf {\bibinfo {volume} {459}},\ \bibinfo {pages}
  {414–418} (\bibinfo {year} {2009})}\BibitemShut {NoStop}%
\bibitem [{\citenamefont {Huang}\ \emph {et~al.}(2016)\citenamefont {Huang},
  \citenamefont {E.},\ and\ \citenamefont {Baker}}]{Huang2016}%
  \BibitemOpen
  \bibfield  {author} {\bibinfo {author} {\bibfnamefont {P.-S.}\ \bibnamefont
  {Huang}}, \bibinfo {author} {\bibfnamefont {B.~S.}\ \bibnamefont {E.}},\ and\
  \bibinfo {author} {\bibfnamefont {D.}~\bibnamefont {Baker}},\ }\bibfield
  {title} {\bibinfo {title} {The coming of age of {\it de novo} protein
  design},\ }\href {https://doi.org/10.1038/nature19946} {\bibfield  {journal}
  {\bibinfo  {journal} {Nature}\ }\textbf {\bibinfo {volume} {537}},\ \bibinfo
  {pages} {320} (\bibinfo {year} {2016})}\BibitemShut {NoStop}%
\bibitem [{\citenamefont {Glotzer}\ and\ \citenamefont
  {Solomon}(2007)}]{Glotzer2007}%
  \BibitemOpen
  \bibfield  {author} {\bibinfo {author} {\bibfnamefont {S.~C.}\ \bibnamefont
  {Glotzer}}\ and\ \bibinfo {author} {\bibfnamefont {M.~J.}\ \bibnamefont
  {Solomon}},\ }\bibfield  {title} {\bibinfo {title} {{Anisotropy of building
  blocks and their assembly into complex structures}},\ }\bibfield  {journal}
  {\bibinfo  {journal} {Nature Materials}\ }\href
  {https://doi.org/10.1038/nmat1949} {10.1038/nmat1949} (\bibinfo {year}
  {2007})\BibitemShut {NoStop}%
\bibitem [{\citenamefont {Sacanna}\ and\ \citenamefont
  {Pine}(2011)}]{Sacanna2011}%
  \BibitemOpen
  \bibfield  {author} {\bibinfo {author} {\bibfnamefont {S.}~\bibnamefont
  {Sacanna}}\ and\ \bibinfo {author} {\bibfnamefont {D.~J.}\ \bibnamefont
  {Pine}},\ }\bibfield  {title} {\bibinfo {title} {{Shape-anisotropic colloids:
  Building blocks for complex assemblies}},\ }\href
  {https://doi.org/10.1016/j.cocis.2011.01.003} {\bibfield  {journal} {\bibinfo
   {journal} {Current Opinion in Colloid {\&} Interface Science}\ }\textbf
  {\bibinfo {volume} {16}},\ \bibinfo {pages} {96} (\bibinfo {year}
  {2011})}\BibitemShut {NoStop}%
\bibitem [{\citenamefont {Heuckel}\ \emph {et~al.}(2021)\citenamefont
  {Heuckel}, \citenamefont {Hocky},\ and\ \citenamefont
  {Sacanna}}]{Heuckel2021}%
  \BibitemOpen
  \bibfield  {author} {\bibinfo {author} {\bibfnamefont {T.}~\bibnamefont
  {Heuckel}}, \bibinfo {author} {\bibfnamefont {G.~M.}\ \bibnamefont {Hocky}},\
  and\ \bibinfo {author} {\bibfnamefont {S.}~\bibnamefont {Sacanna}},\
  }\bibfield  {title} {\bibinfo {title} {Total synthesis of colloidal matter},\
  }\href {https://doi.org/10.1038/s41578-021-00323-x} {\bibfield  {journal}
  {\bibinfo  {journal} {Nature Reviews Materials volume}\ }\textbf {\bibinfo
  {volume} {6}},\ \bibinfo {pages} {1053–1069} (\bibinfo {year}
  {2021})}\BibitemShut {NoStop}%
\bibitem [{\citenamefont {Su}\ \emph {et~al.}(2020)\citenamefont {Su},
  \citenamefont {Zhang}, \citenamefont {Yan}, \citenamefont {Guo},
  \citenamefont {Huang}, \citenamefont {Shan}, \citenamefont {Liu},
  \citenamefont {Liu}, \citenamefont {Huang},\ and\ \citenamefont
  {Cheng}}]{Su2020}%
  \BibitemOpen
  \bibfield  {author} {\bibinfo {author} {\bibfnamefont {Z.}~\bibnamefont
  {Su}}, \bibinfo {author} {\bibfnamefont {R.}~\bibnamefont {Zhang}}, \bibinfo
  {author} {\bibfnamefont {X.-Y.}\ \bibnamefont {Yan}}, \bibinfo {author}
  {\bibfnamefont {Q.-Y.}\ \bibnamefont {Guo}}, \bibinfo {author} {\bibfnamefont
  {J.}~\bibnamefont {Huang}}, \bibinfo {author} {\bibfnamefont
  {W.}~\bibnamefont {Shan}}, \bibinfo {author} {\bibfnamefont {Y.}~\bibnamefont
  {Liu}}, \bibinfo {author} {\bibfnamefont {T.}~\bibnamefont {Liu}}, \bibinfo
  {author} {\bibfnamefont {M.}~\bibnamefont {Huang}},\ and\ \bibinfo {author}
  {\bibfnamefont {S.~Z.}\ \bibnamefont {Cheng}},\ }\bibfield  {title} {\bibinfo
  {title} {{The role of architectural engineering in macromolecular
  self-assemblies via non-covalent interactions: A molecular LEGO approach}},\
  }\href {https://doi.org/10.1016/j.progpolymsci.2020.101230} {\bibfield
  {journal} {\bibinfo  {journal} {Progress in Polymer Science}\ }\textbf
  {\bibinfo {volume} {103}},\ \bibinfo {pages} {101230} (\bibinfo {year}
  {2020})}\BibitemShut {NoStop}%
\bibitem [{\citenamefont {Murugan}\ \emph
  {et~al.}(2015{\natexlab{a}})\citenamefont {Murugan}, \citenamefont {Zou},\
  and\ \citenamefont {Brenner}}]{Murugan2015}%
  \BibitemOpen
  \bibfield  {author} {\bibinfo {author} {\bibfnamefont {A.}~\bibnamefont
  {Murugan}}, \bibinfo {author} {\bibfnamefont {J.}~\bibnamefont {Zou}},\ and\
  \bibinfo {author} {\bibfnamefont {M.~P.}\ \bibnamefont {Brenner}},\
  }\bibfield  {title} {\bibinfo {title} {{Undesired usage and the robust
  self-assembly of heterogeneous structures}},\ }\href
  {https://doi.org/10.1038/ncomms7203} {\bibfield  {journal} {\bibinfo
  {journal} {Nature Communications}\ }\textbf {\bibinfo {volume} {6}},\
  \bibinfo {pages} {6203} (\bibinfo {year} {2015}{\natexlab{a}})}\BibitemShut
  {NoStop}%
\bibitem [{\citenamefont {Murugan}\ \emph
  {et~al.}(2015{\natexlab{b}})\citenamefont {Murugan}, \citenamefont
  {Zeravcic}, \citenamefont {Brenner},\ and\ \citenamefont
  {Leibler}}]{Murugan2015a}%
  \BibitemOpen
  \bibfield  {author} {\bibinfo {author} {\bibfnamefont {A.}~\bibnamefont
  {Murugan}}, \bibinfo {author} {\bibfnamefont {Z.}~\bibnamefont {Zeravcic}},
  \bibinfo {author} {\bibfnamefont {M.~P.}\ \bibnamefont {Brenner}},\ and\
  \bibinfo {author} {\bibfnamefont {S.}~\bibnamefont {Leibler}},\ }\bibfield
  {title} {\bibinfo {title} {{Multifarious assembly mixtures: Systems allowing
  retrieval of diverse stored structures}},\ }\href
  {https://doi.org/10.1073/pnas.1413941112} {\bibfield  {journal} {\bibinfo
  {journal} {Proceedings of the National Academy of Sciences}\ }\textbf
  {\bibinfo {volume} {112}},\ \bibinfo {pages} {54} (\bibinfo {year}
  {2015}{\natexlab{b}})},\ \Eprint {https://arxiv.org/abs/1408.6893}
  {arXiv:1408.6893} \BibitemShut {NoStop}%
\bibitem [{\citenamefont {Jacobs}\ \emph {et~al.}(2015)\citenamefont {Jacobs},
  \citenamefont {Reinhardt},\ and\ \citenamefont {Frenkel}}]{Jacobs2015}%
  \BibitemOpen
  \bibfield  {author} {\bibinfo {author} {\bibfnamefont {W.~M.}\ \bibnamefont
  {Jacobs}}, \bibinfo {author} {\bibfnamefont {A.}~\bibnamefont {Reinhardt}},\
  and\ \bibinfo {author} {\bibfnamefont {D.}~\bibnamefont {Frenkel}},\
  }\bibfield  {title} {\bibinfo {title} {{Rational design of self-assembly
  pathways for complex multicomponent structures}},\ }\href
  {https://doi.org/10.1073/pnas.1502210112} {\bibfield  {journal} {\bibinfo
  {journal} {Proceedings of the National Academy of Sciences}\ }\textbf
  {\bibinfo {volume} {112}},\ \bibinfo {pages} {6313} (\bibinfo {year}
  {2015})},\ \Eprint {https://arxiv.org/abs/arXiv:1502.01351v1}
  {arXiv:arXiv:1502.01351v1} \BibitemShut {NoStop}%
\bibitem [{\citenamefont {Jacobs}\ and\ \citenamefont
  {Frenkel}(2016)}]{Jacobs2016}%
  \BibitemOpen
  \bibfield  {author} {\bibinfo {author} {\bibfnamefont {W.~M.}\ \bibnamefont
  {Jacobs}}\ and\ \bibinfo {author} {\bibfnamefont {D.}~\bibnamefont
  {Frenkel}},\ }\bibfield  {title} {\bibinfo {title} {Self-assembly of
  structures with addressable complexity},\ }\href
  {https://doi.org/10.1021/jacs.5b11918} {\bibfield  {journal} {\bibinfo
  {journal} {Journal of the American Chemical Society}\ }\textbf {\bibinfo
  {volume} {138}},\ \bibinfo {pages} {2457} (\bibinfo {year} {2016})},\
  \bibinfo {note} {pMID: 26862684},\ \Eprint
  {https://arxiv.org/abs/https://doi.org/10.1021/jacs.5b11918}
  {https://doi.org/10.1021/jacs.5b11918} \BibitemShut {NoStop}%
\bibitem [{\citenamefont {Zeravcic}\ \emph {et~al.}(2014)\citenamefont
  {Zeravcic}, \citenamefont {Manoharan},\ and\ \citenamefont
  {Brenner}}]{Zeravcic2014}%
  \BibitemOpen
  \bibfield  {author} {\bibinfo {author} {\bibfnamefont {Z.}~\bibnamefont
  {Zeravcic}}, \bibinfo {author} {\bibfnamefont {V.~N.}\ \bibnamefont
  {Manoharan}},\ and\ \bibinfo {author} {\bibfnamefont {M.~P.}\ \bibnamefont
  {Brenner}},\ }\bibfield  {title} {\bibinfo {title} {{Size limits of
  self-assembled colloidal structures made using specific interactions}},\
  }\href {https://doi.org/10.1073/pnas.1411765111} {\bibfield  {journal}
  {\bibinfo  {journal} {Proceedings of the National Academy of Sciences}\
  }\textbf {\bibinfo {volume} {111}},\ \bibinfo {pages} {15918} (\bibinfo
  {year} {2014})}\BibitemShut {NoStop}%
\bibitem [{\citenamefont {Ke}\ \emph {et~al.}(2014)\citenamefont {Ke},
  \citenamefont {Ong}, \citenamefont {Sun}, \citenamefont {Song}, \citenamefont
  {Dong}, \citenamefont {Shih},\ and\ \citenamefont {Yin}}]{Ke2014}%
  \BibitemOpen
  \bibfield  {author} {\bibinfo {author} {\bibfnamefont {Y.}~\bibnamefont
  {Ke}}, \bibinfo {author} {\bibfnamefont {L.~L.}\ \bibnamefont {Ong}},
  \bibinfo {author} {\bibfnamefont {W.}~\bibnamefont {Sun}}, \bibinfo {author}
  {\bibfnamefont {J.}~\bibnamefont {Song}}, \bibinfo {author} {\bibfnamefont
  {M.}~\bibnamefont {Dong}}, \bibinfo {author} {\bibfnamefont {W.~M.}\
  \bibnamefont {Shih}},\ and\ \bibinfo {author} {\bibfnamefont
  {P.}~\bibnamefont {Yin}},\ }\bibfield  {title} {\bibinfo {title} {{DNA brick
  crystals with prescribed depths}},\ }\href
  {https://doi.org/10.1038/nchem.2083} {\bibfield  {journal} {\bibinfo
  {journal} {Nature Chemistry}\ }\textbf {\bibinfo {volume} {6}},\ \bibinfo
  {pages} {994} (\bibinfo {year} {2014})}\BibitemShut {NoStop}%
\bibitem [{\citenamefont {Ong}\ \emph {et~al.}(2017)\citenamefont {Ong},
  \citenamefont {Hanikel}, \citenamefont {Yaghi}, \citenamefont {Grun},
  \citenamefont {Strauss}, \citenamefont {Bron}, \citenamefont {Lai-Kee-Him},
  \citenamefont {Schueder}, \citenamefont {Wang}, \citenamefont {Wang},
  \citenamefont {Kishi}, \citenamefont {Myhrvold}, \citenamefont {Zhu},
  \citenamefont {Jungmann}, \citenamefont {Bellot}, \citenamefont {Ke},\ and\
  \citenamefont {Yin}}]{Ong2017}%
  \BibitemOpen
  \bibfield  {author} {\bibinfo {author} {\bibfnamefont {L.~L.}\ \bibnamefont
  {Ong}}, \bibinfo {author} {\bibfnamefont {N.}~\bibnamefont {Hanikel}},
  \bibinfo {author} {\bibfnamefont {O.~K.}\ \bibnamefont {Yaghi}}, \bibinfo
  {author} {\bibfnamefont {C.}~\bibnamefont {Grun}}, \bibinfo {author}
  {\bibfnamefont {M.~T.}\ \bibnamefont {Strauss}}, \bibinfo {author}
  {\bibfnamefont {P.}~\bibnamefont {Bron}}, \bibinfo {author} {\bibfnamefont
  {J.}~\bibnamefont {Lai-Kee-Him}}, \bibinfo {author} {\bibfnamefont
  {F.}~\bibnamefont {Schueder}}, \bibinfo {author} {\bibfnamefont
  {B.}~\bibnamefont {Wang}}, \bibinfo {author} {\bibfnamefont {P.}~\bibnamefont
  {Wang}}, \bibinfo {author} {\bibfnamefont {J.~Y.}\ \bibnamefont {Kishi}},
  \bibinfo {author} {\bibfnamefont {C.}~\bibnamefont {Myhrvold}}, \bibinfo
  {author} {\bibfnamefont {A.}~\bibnamefont {Zhu}}, \bibinfo {author}
  {\bibfnamefont {R.}~\bibnamefont {Jungmann}}, \bibinfo {author}
  {\bibfnamefont {G.}~\bibnamefont {Bellot}}, \bibinfo {author} {\bibfnamefont
  {Y.}~\bibnamefont {Ke}},\ and\ \bibinfo {author} {\bibfnamefont
  {P.}~\bibnamefont {Yin}},\ }\bibfield  {title} {\bibinfo {title}
  {{Programmable self-assembly of three-dimensional nanostructures from 10,000
  unique components}},\ }\href {https://doi.org/10.1038/nature24648} {\bibfield
   {journal} {\bibinfo  {journal} {Nature}\ }\textbf {\bibinfo {volume}
  {552}},\ \bibinfo {pages} {72} (\bibinfo {year} {2017})}\BibitemShut
  {NoStop}%
\bibitem [{\citenamefont {Nykypanchuk}\ \emph {et~al.}(2008)\citenamefont
  {Nykypanchuk}, \citenamefont {Maye}, \citenamefont {Van Der~Lelie},\ and\
  \citenamefont {Gang}}]{Nykypanchuk2008}%
  \BibitemOpen
  \bibfield  {author} {\bibinfo {author} {\bibfnamefont {D.}~\bibnamefont
  {Nykypanchuk}}, \bibinfo {author} {\bibfnamefont {M.~M.}\ \bibnamefont
  {Maye}}, \bibinfo {author} {\bibfnamefont {D.}~\bibnamefont {Van
  Der~Lelie}},\ and\ \bibinfo {author} {\bibfnamefont {O.}~\bibnamefont
  {Gang}},\ }\bibfield  {title} {{\bibinfo {title}
  {{DNA}-guided crystallization of colloidal nanoparticles}},\ }\href
  {https://doi.org/10.1038/nature06560} {\bibfield  {journal} {\bibinfo
  {journal} {Nature}\ }\textbf {\bibinfo {volume} {451}},\ \bibinfo {pages}
  {549} (\bibinfo {year} {2008})}\BibitemShut {NoStop}%
\bibitem [{\citenamefont {Auyeung}\ \emph {et~al.}(2012)\citenamefont
  {Auyeung}, \citenamefont {Cutler}, \citenamefont {Macfarlane}, \citenamefont
  {Jones}, \citenamefont {Wu}, \citenamefont {Liu}, \citenamefont {Zhang},
  \citenamefont {Osberg},\ and\ \citenamefont {Mirkin}}]{Auyeung2012}%
  \BibitemOpen
  \bibfield  {author} {\bibinfo {author} {\bibfnamefont {E.}~\bibnamefont
  {Auyeung}}, \bibinfo {author} {\bibfnamefont {J.~I.}\ \bibnamefont {Cutler}},
  \bibinfo {author} {\bibfnamefont {R.~J.}\ \bibnamefont {Macfarlane}},
  \bibinfo {author} {\bibfnamefont {M.~R.}\ \bibnamefont {Jones}}, \bibinfo
  {author} {\bibfnamefont {J.}~\bibnamefont {Wu}}, \bibinfo {author}
  {\bibfnamefont {G.}~\bibnamefont {Liu}}, \bibinfo {author} {\bibfnamefont
  {K.}~\bibnamefont {Zhang}}, \bibinfo {author} {\bibfnamefont {K.~D.}\
  \bibnamefont {Osberg}},\ and\ \bibinfo {author} {\bibfnamefont {C.~A.}\
  \bibnamefont {Mirkin}},\ }\bibfield  {title} {\bibinfo {title}
  {{Synthetically programmable nanoparticle superlattices using a hollow
  three-dimensional spacer approach}},\ }\href
  {https://doi.org/10.1038/nnano.2011.222} {\bibfield  {journal} {\bibinfo
  {journal} {Nature Nanotechnology}\ }\textbf {\bibinfo {volume} {7}},\
  \bibinfo {pages} {24} (\bibinfo {year} {2012})}\BibitemShut {NoStop}%
\bibitem [{\citenamefont {Tian}\ \emph {et~al.}(2020)\citenamefont {Tian},
  \citenamefont {Lhermitte}, \citenamefont {Bai}, \citenamefont {Vo},
  \citenamefont {L.~Xin}, \citenamefont {Li}, \citenamefont {Li}, \citenamefont
  {Fukuto}, \citenamefont {Yager}, \citenamefont {Kahn}, \citenamefont {Xiong},
  \citenamefont {Minevich},\ and\ \citenamefont {Gang}}]{Tian2020}%
  \BibitemOpen
  \bibfield  {author} {\bibinfo {author} {\bibfnamefont {Y.}~\bibnamefont
  {Tian}}, \bibinfo {author} {\bibfnamefont {J.~R.}\ \bibnamefont {Lhermitte}},
  \bibinfo {author} {\bibfnamefont {L.}~\bibnamefont {Bai}}, \bibinfo {author}
  {\bibfnamefont {T.}~\bibnamefont {Vo}}, \bibinfo {author} {\bibfnamefont
  {H.}~\bibnamefont {L.~Xin}}, \bibinfo {author} {\bibfnamefont
  {H.}~\bibnamefont {Li}}, \bibinfo {author} {\bibfnamefont {R.}~\bibnamefont
  {Li}}, \bibinfo {author} {\bibfnamefont {M.}~\bibnamefont {Fukuto}}, \bibinfo
  {author} {\bibfnamefont {K.~G.}\ \bibnamefont {Yager}}, \bibinfo {author}
  {\bibfnamefont {J.~S.}\ \bibnamefont {Kahn}}, \bibinfo {author}
  {\bibfnamefont {Y.}~\bibnamefont {Xiong}}, \bibinfo {author} {\bibfnamefont
  {S.~K.}\ \bibnamefont {Minevich}, \bibfnamefont {Brian an~Kumar}},\ and\
  \bibinfo {author} {\bibfnamefont {O.}~\bibnamefont {Gang}},\ }\bibfield
  {title} {\bibinfo {title} {Ordered three-dimensional nanomaterials using
  dna-prescribed and valence-controlled material voxels},\ }\href@noop {}
  {\bibfield  {journal} {\bibinfo  {journal} {Nature Materials}\ }\textbf
  {\bibinfo {volume} {19}},\ \bibinfo {pages} {789–796} (\bibinfo {year}
  {2020})}\BibitemShut {NoStop}%
\bibitem [{\citenamefont {Rogers}\ and\ \citenamefont
  {Crocker}(2011)}]{Rogers2011}%
  \BibitemOpen
  \bibfield  {author} {\bibinfo {author} {\bibfnamefont {W.~B.}\ \bibnamefont
  {Rogers}}\ and\ \bibinfo {author} {\bibfnamefont {J.~C.}\ \bibnamefont
  {Crocker}},\ }\bibfield  {title} {\bibinfo {title} {Direct measurements of
  dna-mediated colloidal interactions and their quantitative modeling},\ }\href
  {https://doi.org/10.1073/pnas.1109853108} {\bibfield  {journal} {\bibinfo
  {journal} {Proceedings of the National Academy of Sciences}\ }\textbf
  {\bibinfo {volume} {108}},\ \bibinfo {pages} {15687} (\bibinfo {year}
  {2011})},\ \Eprint
  {https://arxiv.org/abs/https://www.pnas.org/doi/pdf/10.1073/pnas.1109853108}
  {https://www.pnas.org/doi/pdf/10.1073/pnas.1109853108} \BibitemShut {NoStop}%
\bibitem [{\citenamefont {Hensley}\ \emph {et~al.}(2022)\citenamefont
  {Hensley}, \citenamefont {Jacobs},\ and\ \citenamefont
  {Rogers}}]{Hensley2022}%
  \BibitemOpen
  \bibfield  {author} {\bibinfo {author} {\bibfnamefont {A.}~\bibnamefont
  {Hensley}}, \bibinfo {author} {\bibfnamefont {W.~M.}\ \bibnamefont
  {Jacobs}},\ and\ \bibinfo {author} {\bibfnamefont {W.~B.}\ \bibnamefont
  {Rogers}},\ }\bibfield  {title} {\bibinfo {title} {Self-assembly of photonic
  crystals by controlling the nucleation and growth of dna-coated colloids},\
  }\href {https://doi.org/10.1073/pnas.2114050118} {\bibfield  {journal}
  {\bibinfo  {journal} {Proceedings of the National Academy of Sciences}\
  }\textbf {\bibinfo {volume} {119}},\ \bibinfo {pages} {e2114050118} (\bibinfo
  {year} {2022})},\ \Eprint
  {https://arxiv.org/abs/https://www.pnas.org/doi/pdf/10.1073/pnas.2114050118}
  {https://www.pnas.org/doi/pdf/10.1073/pnas.2114050118} \BibitemShut {NoStop}%
\bibitem [{\citenamefont {Majewski}\ \emph {et~al.}(2021)\citenamefont
  {Majewski}, \citenamefont {Michelson}, \citenamefont {Cordeiro},
  \citenamefont {Tian}, \citenamefont {Ma}, \citenamefont {Kisslinger},
  \citenamefont {Tian}, \citenamefont {Liu}, \citenamefont {Stach},
  \citenamefont {Yager},\ and\ \citenamefont {Gang}}]{Majewski2021}%
  \BibitemOpen
  \bibfield  {author} {\bibinfo {author} {\bibfnamefont {P.~W.}\ \bibnamefont
  {Majewski}}, \bibinfo {author} {\bibfnamefont {A.}~\bibnamefont {Michelson}},
  \bibinfo {author} {\bibfnamefont {M.~A.~L.}\ \bibnamefont {Cordeiro}},
  \bibinfo {author} {\bibfnamefont {C.}~\bibnamefont {Tian}}, \bibinfo {author}
  {\bibfnamefont {C.}~\bibnamefont {Ma}}, \bibinfo {author} {\bibfnamefont
  {K.}~\bibnamefont {Kisslinger}}, \bibinfo {author} {\bibfnamefont
  {Y.}~\bibnamefont {Tian}}, \bibinfo {author} {\bibfnamefont {W.}~\bibnamefont
  {Liu}}, \bibinfo {author} {\bibfnamefont {E.~A.}\ \bibnamefont {Stach}},
  \bibinfo {author} {\bibfnamefont {K.~G.}\ \bibnamefont {Yager}},\ and\
  \bibinfo {author} {\bibfnamefont {O.}~\bibnamefont {Gang}},\ }\bibfield
  {title} {\bibinfo {title} {Resilient three-dimensional ordered architectures
  assembled from nanoparticles by dna},\ }\href
  {https://doi.org/10.1126/sciadv.abf0617} {\bibfield  {journal} {\bibinfo
  {journal} {Science Advances}\ }\textbf {\bibinfo {volume} {7}},\ \bibinfo
  {pages} {eabf0617} (\bibinfo {year} {2021})},\ \Eprint
  {https://arxiv.org/abs/https://www.science.org/doi/pdf/10.1126/sciadv.abf0617}
  {https://www.science.org/doi/pdf/10.1126/sciadv.abf0617} \BibitemShut
  {NoStop}%
\bibitem [{\citenamefont {Michelson}\ \emph {et~al.}(2022)\citenamefont
  {Michelson}, \citenamefont {Minevich}, \citenamefont {Emamy}, \citenamefont
  {Huang}, \citenamefont {Chu}, \citenamefont {Yan},\ and\ \citenamefont
  {Gang}}]{Michelson2022}%
  \BibitemOpen
  \bibfield  {author} {\bibinfo {author} {\bibfnamefont {A.}~\bibnamefont
  {Michelson}}, \bibinfo {author} {\bibfnamefont {B.}~\bibnamefont {Minevich}},
  \bibinfo {author} {\bibfnamefont {H.}~\bibnamefont {Emamy}}, \bibinfo
  {author} {\bibfnamefont {X.}~\bibnamefont {Huang}}, \bibinfo {author}
  {\bibfnamefont {Y.~S.}\ \bibnamefont {Chu}}, \bibinfo {author} {\bibfnamefont
  {H.}~\bibnamefont {Yan}},\ and\ \bibinfo {author} {\bibfnamefont
  {O.}~\bibnamefont {Gang}},\ }\bibfield  {title} {\bibinfo {title}
  {Three-dimensional visualization of nanoparticle lattices and multimaterial
  frameworks},\ }\href {https://doi.org/10.1126/science.abk0463} {\bibfield
  {journal} {\bibinfo  {journal} {Science}\ }\textbf {\bibinfo {volume}
  {376}},\ \bibinfo {pages} {203} (\bibinfo {year} {2022})},\ \Eprint
  {https://arxiv.org/abs/https://www.science.org/doi/pdf/10.1126/science.abk0463}
  {https://www.science.org/doi/pdf/10.1126/science.abk0463} \BibitemShut
  {NoStop}%
\bibitem [{\citenamefont {Park}\ \emph {et~al.}(2020)\citenamefont {Park},
  \citenamefont {Park}, \citenamefont {Hur},\ and\ \citenamefont
  {Lee}}]{Park2020}%
  \BibitemOpen
  \bibfield  {author} {\bibinfo {author} {\bibfnamefont {S.~H.}\ \bibnamefont
  {Park}}, \bibinfo {author} {\bibfnamefont {H.}~\bibnamefont {Park}}, \bibinfo
  {author} {\bibfnamefont {K.}~\bibnamefont {Hur}},\ and\ \bibinfo {author}
  {\bibfnamefont {S.}~\bibnamefont {Lee}},\ }\bibfield  {title} {\bibinfo
  {title} {Design of dna origami diamond photonic crystals},\ }\href
  {https://doi.org/10.1021/acsabm.9b01171} {\bibfield  {journal} {\bibinfo
  {journal} {ACS Applied Bio Materials}\ }\textbf {\bibinfo {volume} {3}},\
  \bibinfo {pages} {747} (\bibinfo {year} {2020})},\ \bibinfo {note} {pMID:
  35019418},\ \Eprint
  {https://arxiv.org/abs/https://doi.org/10.1021/acsabm.9b01171}
  {https://doi.org/10.1021/acsabm.9b01171} \BibitemShut {NoStop}%
\bibitem [{\citenamefont {Gerling}\ \emph {et~al.}(2015)\citenamefont
  {Gerling}, \citenamefont {Wagenbauer}, \citenamefont {Neuner},\ and\
  \citenamefont {Dietz}}]{Gerling2015}%
  \BibitemOpen
  \bibfield  {author} {\bibinfo {author} {\bibfnamefont {T.}~\bibnamefont
  {Gerling}}, \bibinfo {author} {\bibfnamefont {K.~F.}\ \bibnamefont
  {Wagenbauer}}, \bibinfo {author} {\bibfnamefont {A.~M.}\ \bibnamefont
  {Neuner}},\ and\ \bibinfo {author} {\bibfnamefont {H.}~\bibnamefont
  {Dietz}},\ }\bibfield  {title} {\bibinfo {title} {Dynamic dna devices and
  assemblies formed by shape-complementary, non-base pairing 3d
  components},\ }\href {https://doi.org/10.1126/science.aaa5372} {\bibfield
  {journal} {\bibinfo  {journal} {Science}\ }\textbf {\bibinfo {volume}
  {347}},\ \bibinfo {pages} {1446} (\bibinfo {year} {2015})},\ \Eprint
  {https://arxiv.org/abs/https://www.science.org/doi/pdf/10.1126/science.aaa5372}
  {https://www.science.org/doi/pdf/10.1126/science.aaa5372} \BibitemShut
  {NoStop}%
\bibitem [{\citenamefont {Rothemund}\ \emph {et~al.}(2004)\citenamefont
  {Rothemund}, \citenamefont {Ekani-Nkodo}, \citenamefont {Papadakis},
  \citenamefont {Kumar}, \citenamefont {Fygenson},\ and\ \citenamefont
  {Winfree}}]{Rothemund2004}%
  \BibitemOpen
  \bibfield  {author} {\bibinfo {author} {\bibfnamefont {P.~W.~K.}\
  \bibnamefont {Rothemund}}, \bibinfo {author} {\bibfnamefont {A.}~\bibnamefont
  {Ekani-Nkodo}}, \bibinfo {author} {\bibfnamefont {N.}~\bibnamefont
  {Papadakis}}, \bibinfo {author} {\bibfnamefont {A.}~\bibnamefont {Kumar}},
  \bibinfo {author} {\bibfnamefont {D.~K.}\ \bibnamefont {Fygenson}},\ and\
  \bibinfo {author} {\bibfnamefont {E.}~\bibnamefont {Winfree}},\ }\bibfield
  {title} {\bibinfo {title} {Design and characterization of programmable dna
  nanotubes},\ }\href {https://doi.org/10.1021/ja044319l} {\bibfield  {journal}
  {\bibinfo  {journal} {Journal of the American Chemical Society}\ }\textbf
  {\bibinfo {volume} {126}},\ \bibinfo {pages} {16344} (\bibinfo {year}
  {2004})},\ \bibinfo {note} {pMID: 15600335},\ \Eprint
  {https://arxiv.org/abs/https://doi.org/10.1021/ja044319l}
  {https://doi.org/10.1021/ja044319l} \BibitemShut {NoStop}%
\bibitem [{\citenamefont {Benson}\ \emph {et~al.}(2015)\citenamefont {Benson},
  \citenamefont {Mohammed}, \citenamefont {Gardell}, \citenamefont {Masich},
  \citenamefont {Czeizler}, \citenamefont {Orponen},\ and\ \citenamefont
  {H\"ogberg}}]{Benson2015}%
  \BibitemOpen
  \bibfield  {author} {\bibinfo {author} {\bibfnamefont {E.}~\bibnamefont
  {Benson}}, \bibinfo {author} {\bibfnamefont {A.}~\bibnamefont {Mohammed}},
  \bibinfo {author} {\bibfnamefont {J.}~\bibnamefont {Gardell}}, \bibinfo
  {author} {\bibfnamefont {S.}~\bibnamefont {Masich}}, \bibinfo {author}
  {\bibfnamefont {E.}~\bibnamefont {Czeizler}}, \bibinfo {author}
  {\bibfnamefont {P.}~\bibnamefont {Orponen}},\ and\ \bibinfo {author}
  {\bibfnamefont {B.}~\bibnamefont {H\"ogberg}},\ }\bibfield  {title} {\bibinfo
  {title} {Dna rendering of polyhedral meshes at the nanoscale},\ }\href@noop
  {} {\bibfield  {journal} {\bibinfo  {journal} {Nature}\ }\textbf {\bibinfo
  {volume} {523}},\ \bibinfo {pages} {441–444} (\bibinfo {year}
  {2015})}\BibitemShut {NoStop}%
\bibitem [{\citenamefont {Sigl}\ \emph {et~al.}(2021)\citenamefont {Sigl},
  \citenamefont {Willne}, \citenamefont {Engelen}, \citenamefont {Kretzmann},
  \citenamefont {Sachenbacher}, \citenamefont {Liedl}, \citenamefont {Kolbe},
  \citenamefont {Wilsch}, \citenamefont {Aghvami}, \citenamefont {Protzer},
  \citenamefont {Hagan}, \citenamefont {Fraden},\ and\ \citenamefont
  {Dietz}}]{Sigl2021}%
  \BibitemOpen
  \bibfield  {author} {\bibinfo {author} {\bibfnamefont {C.}~\bibnamefont
  {Sigl}}, \bibinfo {author} {\bibfnamefont {E.~M.}\ \bibnamefont {Willne}},
  \bibinfo {author} {\bibfnamefont {W.}~\bibnamefont {Engelen}}, \bibinfo
  {author} {\bibfnamefont {J.~A.}\ \bibnamefont {Kretzmann}}, \bibinfo {author}
  {\bibfnamefont {K.}~\bibnamefont {Sachenbacher}}, \bibinfo {author}
  {\bibfnamefont {A.}~\bibnamefont {Liedl}}, \bibinfo {author} {\bibfnamefont
  {F.}~\bibnamefont {Kolbe}}, \bibinfo {author} {\bibfnamefont
  {F.}~\bibnamefont {Wilsch}}, \bibinfo {author} {\bibfnamefont {S.~A.}\
  \bibnamefont {Aghvami}}, \bibinfo {author} {\bibfnamefont {U.}~\bibnamefont
  {Protzer}}, \bibinfo {author} {\bibfnamefont {M.~F.}\ \bibnamefont {Hagan}},
  \bibinfo {author} {\bibfnamefont {S.}~\bibnamefont {Fraden}},\ and\ \bibinfo
  {author} {\bibfnamefont {H.}~\bibnamefont {Dietz}},\ }\bibfield  {title}
  {\bibinfo {title} {Programmable icosahedral shell system for virus
  trapping},\ }\href@noop {} {\bibfield  {journal} {\bibinfo  {journal} {Nature
  Materials}\ }\textbf {\bibinfo {volume} {20}},\ \bibinfo {pages}
  {1281–1289} (\bibinfo {year} {2021})}\BibitemShut {NoStop}%
\bibitem [{\citenamefont {Videbaek}\ \emph {et~al.}(2022)\citenamefont
  {Videbaek}, \citenamefont {Fang}, \citenamefont {Hayakawa}, \citenamefont
  {Tyukodi}, \citenamefont {Hagan},\ and\ \citenamefont
  {Rogers}}]{Videbaek2022}%
  \BibitemOpen
  \bibfield  {author} {\bibinfo {author} {\bibfnamefont {T.~E.}\ \bibnamefont
  {Videbaek}}, \bibinfo {author} {\bibfnamefont {H.}~\bibnamefont {Fang}},
  \bibinfo {author} {\bibfnamefont {D.}~\bibnamefont {Hayakawa}}, \bibinfo
  {author} {\bibfnamefont {B.}~\bibnamefont {Tyukodi}}, \bibinfo {author}
  {\bibfnamefont {M.~F.}\ \bibnamefont {Hagan}},\ and\ \bibinfo {author}
  {\bibfnamefont {W.~B.}\ \bibnamefont {Rogers}},\ }\bibfield  {title}
  {\bibinfo {title} {Tiling a tubule: how increasing complexity improves the
  yield of self-limited assembly},\ }\href
  {https://doi.org/10.1088/1361-648X/ac47dd} {\bibfield  {journal} {\bibinfo
  {journal} {Journal of Physics: Condensed Matter}\ }\textbf {\bibinfo {volume}
  {34}},\ \bibinfo {pages} {134003} (\bibinfo {year} {2022})}\BibitemShut
  {NoStop}%
\bibitem [{\citenamefont {Hayakawa}\ \emph {et~al.}(2022)\citenamefont
  {Hayakawa}, \citenamefont {Videbaek}, \citenamefont {Hall}, \citenamefont
  {Fang}, \citenamefont {Sigl}, \citenamefont {Feigl}, \citenamefont {Dietz},
  \citenamefont {Fraden}, \citenamefont {Hagan}, \citenamefont {Grason},\ and\
  \citenamefont {Rogers}}]{Hayakawa2022}%
  \BibitemOpen
  \bibfield  {author} {\bibinfo {author} {\bibfnamefont {D.}~\bibnamefont
  {Hayakawa}}, \bibinfo {author} {\bibfnamefont {T.~E.}\ \bibnamefont
  {Videbaek}}, \bibinfo {author} {\bibfnamefont {D.~M.}\ \bibnamefont {Hall}},
  \bibinfo {author} {\bibfnamefont {H.}~\bibnamefont {Fang}}, \bibinfo {author}
  {\bibfnamefont {C.}~\bibnamefont {Sigl}}, \bibinfo {author} {\bibfnamefont
  {E.}~\bibnamefont {Feigl}}, \bibinfo {author} {\bibfnamefont
  {H.}~\bibnamefont {Dietz}}, \bibinfo {author} {\bibfnamefont
  {S.}~\bibnamefont {Fraden}}, \bibinfo {author} {\bibfnamefont {M.~F.}\
  \bibnamefont {Hagan}}, \bibinfo {author} {\bibfnamefont {G.~M.}\ \bibnamefont
  {Grason}},\ and\ \bibinfo {author} {\bibfnamefont {W.~B.}\ \bibnamefont
  {Rogers}},\ }\bibfield  {title} {\bibinfo {title} {Geometrically programmed
  self-limited assembly of tubules using dna origami colloids},\ }\href
  {https://doi.org/10.1073/pnas.2207902119} {\bibfield  {journal} {\bibinfo
  {journal} {Proceedings of the National Academy of Sciences}\ }\textbf
  {\bibinfo {volume} {119}},\ \bibinfo {pages} {e2207902119} (\bibinfo {year}
  {2022})},\ \Eprint
  {https://arxiv.org/abs/https://www.pnas.org/doi/pdf/10.1073/pnas.2207902119}
  {https://www.pnas.org/doi/pdf/10.1073/pnas.2207902119} \BibitemShut {NoStop}%
\bibitem [{\citenamefont {Zandi}\ \emph {et~al.}(2004)\citenamefont {Zandi},
  \citenamefont {Reguera}, \citenamefont {Bruinsma}, \citenamefont {Gelbart},\
  and\ \citenamefont {Rudnick}}]{Zandi2004}%
  \BibitemOpen
  \bibfield  {author} {\bibinfo {author} {\bibfnamefont {R.}~\bibnamefont
  {Zandi}}, \bibinfo {author} {\bibfnamefont {D.}~\bibnamefont {Reguera}},
  \bibinfo {author} {\bibfnamefont {R.~F.}\ \bibnamefont {Bruinsma}}, \bibinfo
  {author} {\bibfnamefont {W.~M.}\ \bibnamefont {Gelbart}},\ and\ \bibinfo
  {author} {\bibfnamefont {J.}~\bibnamefont {Rudnick}},\ }\bibfield  {title}
  {\bibinfo {title} {Origin of icosahedral symmetry in viruses},\ }\href
  {https://doi.org/10.1073/pnas.0405844101} {\bibfield  {journal} {\bibinfo
  {journal} {Proceedings of the National Academy of Sciences}\ }\textbf
  {\bibinfo {volume} {101}},\ \bibinfo {pages} {15556} (\bibinfo {year}
  {2004})},\ \Eprint
  {https://arxiv.org/abs/https://www.pnas.org/doi/pdf/10.1073/pnas.0405844101}
  {https://www.pnas.org/doi/pdf/10.1073/pnas.0405844101} \BibitemShut {NoStop}%
\bibitem [{\citenamefont {Twarock}\ and\ \citenamefont
  {Luque}(2019)}]{Twarock2019}%
  \BibitemOpen
  \bibfield  {author} {\bibinfo {author} {\bibfnamefont {R.}~\bibnamefont
  {Twarock}}\ and\ \bibinfo {author} {\bibfnamefont {A.}~\bibnamefont
  {Luque}},\ }\bibfield  {title} {{\bibinfo {title}
  {Structural puzzles in virology solved with an overarching icosahedral design
  principle}},\ }\href {https://doi.org/10.1038/s41467-019-12367-3} {\bibfield
  {journal} {\bibinfo  {journal} {Nature Communications}\ }\textbf {\bibinfo
  {volume} {10}},\ \bibinfo {pages} {4414} (\bibinfo {year}
  {2019})}\BibitemShut {NoStop}%
\bibitem [{\citenamefont {Siber}(2020)}]{Siber2020}%
  \BibitemOpen
  \bibfield  {author} {\bibinfo {author} {\bibfnamefont {A.}~\bibnamefont
  {Siber}},\ }\bibfield  {title} {\bibinfo {title} {Icosadeltahedral geometry
  of geodesic domes, fullerenes and viruses: A tutorial on the t-number},\
  }\bibfield  {journal} {\bibinfo  {journal} {Symmetry}\ }\textbf {\bibinfo
  {volume} {12}},\ \href {https://doi.org/10.3390/sym12040556}
  {10.3390/sym12040556} (\bibinfo {year} {2020})\BibitemShut {NoStop}%
\bibitem [{\citenamefont {Johnson}\ and\ \citenamefont
  {Speir}(1997)}]{Johnson1997}%
  \BibitemOpen
  \bibfield  {author} {\bibinfo {author} {\bibfnamefont {J.~E.}\ \bibnamefont
  {Johnson}}\ and\ \bibinfo {author} {\bibfnamefont {J.~A.}\ \bibnamefont
  {Speir}},\ }\bibfield  {title} {\bibinfo {title} {{Quasi-equivalent viruses:
  A paradigm for protein assemblies}},\ }\href@noop {} {\bibfield  {journal}
  {\bibinfo  {journal} {J. Mol. Biol.}\ }\textbf {\bibinfo {volume} {269}},\
  \bibinfo {pages} {665} (\bibinfo {year} {1997})}\BibitemShut {NoStop}%
\bibitem [{\citenamefont {Caspar}\ and\ \citenamefont
  {Klug}(1962)}]{Caspar1962}%
  \BibitemOpen
  \bibfield  {author} {\bibinfo {author} {\bibfnamefont {D.~L.~D.}\
  \bibnamefont {Caspar}}\ and\ \bibinfo {author} {\bibfnamefont
  {A.}~\bibnamefont {Klug}},\ }\bibfield  {title} {\bibinfo {title} {{Physical
  Principles in the Construction of Regular Viruses}},\ }\href
  {https://doi.org/10.1101/SQB.1962.027.001.005} {\bibfield  {journal}
  {\bibinfo  {journal} {Cold Spring Harbor Symposia on Quantitative Biology}\
  }\textbf {\bibinfo {volume} {27}},\ \bibinfo {pages} {1} (\bibinfo {year}
  {1962})}\BibitemShut {NoStop}%
\bibitem [{\citenamefont {Bohlin}\ \emph {et~al.}(2023)\citenamefont {Bohlin},
  \citenamefont {Turberfield}, \citenamefont {Louis},\ and\ \citenamefont
  {Å ulc}}]{Bohlin2023}%
  \BibitemOpen
  \bibfield  {author} {\bibinfo {author} {\bibfnamefont {J.}~\bibnamefont
  {Bohlin}}, \bibinfo {author} {\bibfnamefont {A.~J.}\ \bibnamefont
  {Turberfield}}, \bibinfo {author} {\bibfnamefont {A.~A.}\ \bibnamefont
  {Louis}},\ and\ \bibinfo {author} {\bibfnamefont {P.}~\bibnamefont
  {Å ulc}},\ }\bibfield  {title} {\bibinfo {title} {Designing the
  self-assembly of arbitrary shapes using minimal complexity building blocks},\
  }\href {https://doi.org/10.1021/acsnano.2c09677} {\bibfield  {journal}
  {\bibinfo  {journal} {ACS Nano}\ }\textbf {\bibinfo {volume} {17}},\ \bibinfo
  {pages} {5387} (\bibinfo {year} {2023})},\ \bibinfo {note} {pMID: 36763807},\
  \Eprint {https://arxiv.org/abs/https://doi.org/10.1021/acsnano.2c09677}
  {https://doi.org/10.1021/acsnano.2c09677} \BibitemShut {NoStop}%
\bibitem [{\citenamefont {Pinto}\ \emph {et~al.}(2023)\citenamefont {Pinto},
  \citenamefont {Sulc}, \citenamefont {Sciortino},\ and\ \citenamefont
  {Russo}}]{Pinto2023}%
  \BibitemOpen
  \bibfield  {author} {\bibinfo {author} {\bibfnamefont {D.~E.~P.}\
  \bibnamefont {Pinto}}, \bibinfo {author} {\bibfnamefont {P.}~\bibnamefont
  {Sulc}}, \bibinfo {author} {\bibfnamefont {F.}~\bibnamefont {Sciortino}},\
  and\ \bibinfo {author} {\bibfnamefont {J.}~\bibnamefont {Russo}},\ }\bibfield
   {title} {\bibinfo {title} {Design strategies for the self-assembly of
  polyhedral shells},\ }\href {https://doi.org/10.1073/pnas.2219458120}
  {\bibfield  {journal} {\bibinfo  {journal} {Proceedings of the National
  Academy of Sciences}\ }\textbf {\bibinfo {volume} {120}},\ \bibinfo {pages}
  {e2219458120} (\bibinfo {year} {2023})},\ \Eprint
  {https://arxiv.org/abs/https://www.pnas.org/doi/pdf/10.1073/pnas.2219458120}
  {https://www.pnas.org/doi/pdf/10.1073/pnas.2219458120} \BibitemShut {NoStop}%
\bibitem [{\citenamefont {Pedersen}\ and\ \citenamefont
  {Hyde}(2018)}]{Pedersen2018}%
  \BibitemOpen
  \bibfield  {author} {\bibinfo {author} {\bibfnamefont {M.~C.}\ \bibnamefont
  {Pedersen}}\ and\ \bibinfo {author} {\bibfnamefont {S.~T.}\ \bibnamefont
  {Hyde}},\ }\bibfield  {title} {\bibinfo {title} {Polyhedra and packings from
  hyperbolic honeycombs},\ }\href {https://doi.org/10.1073/pnas.1720307115}
  {\bibfield  {journal} {\bibinfo  {journal} {Proceedings of the National
  Academy of Sciences}\ }\textbf {\bibinfo {volume} {115}},\ \bibinfo {pages}
  {6905} (\bibinfo {year} {2018})},\ \Eprint
  {https://arxiv.org/abs/https://www.pnas.org/doi/pdf/10.1073/pnas.1720307115}
  {https://www.pnas.org/doi/pdf/10.1073/pnas.1720307115} \BibitemShut {NoStop}%
\bibitem [{\citenamefont {Tanaka}\ \emph {et~al.}(0)\citenamefont {Tanaka},
  \citenamefont {Dotera},\ and\ \citenamefont {Hyde}}]{Tanaka2023}%
  \BibitemOpen
  \bibfield  {author} {\bibinfo {author} {\bibfnamefont {H.}~\bibnamefont
  {Tanaka}}, \bibinfo {author} {\bibfnamefont {T.}~\bibnamefont {Dotera}},\
  and\ \bibinfo {author} {\bibfnamefont {S.~T.}\ \bibnamefont {Hyde}},\
  }\bibfield  {title} {\bibinfo {title} {Programmable self-assembly of
  nanoplates into bicontinuous nanostructures},\ }\href
  {https://doi.org/10.1021/acsnano.2c11929} {\bibfield  {journal} {\bibinfo
  {journal} {ACS Nano}\ }\textbf {\bibinfo {volume} {0}},\ \bibinfo {pages}
  {null} (\bibinfo {year} {0})},\ \bibinfo {note} {pMID: 37527198},\ \Eprint
  {https://arxiv.org/abs/https://doi.org/10.1021/acsnano.2c11929}
  {https://doi.org/10.1021/acsnano.2c11929} \BibitemShut {NoStop}%
\bibitem [{\citenamefont {Lord}\ and\ \citenamefont {Mackay}(2003)}]{Lord2003}%
  \BibitemOpen
  \bibfield  {author} {\bibinfo {author} {\bibfnamefont {E.~A.}\ \bibnamefont
  {Lord}}\ and\ \bibinfo {author} {\bibfnamefont {A.~L.}\ \bibnamefont
  {Mackay}},\ }\bibfield  {title} {\bibinfo {title} {Periodic minimal surfaces
  of cubic symmetry},\ }\href {http://www.jstor.org/stable/24108665} {\bibfield
   {journal} {\bibinfo  {journal} {Current Science}\ }\textbf {\bibinfo
  {volume} {85}},\ \bibinfo {pages} {346} (\bibinfo {year} {2003})}\BibitemShut
  {NoStop}%
\bibitem [{\citenamefont {Schoen}(2012)}]{Schoen2012}%
  \BibitemOpen
  \bibfield  {author} {\bibinfo {author} {\bibfnamefont {A.~H.}\ \bibnamefont
  {Schoen}},\ }\bibfield  {title} {\bibinfo {title} {Reflections concerning
  triply-periodic minimal surfaces},\ }\href
  {https://doi.org/10.1098/rsfs.2012.0023} {\bibfield  {journal} {\bibinfo
  {journal} {Interface Focus}\ }\textbf {\bibinfo {volume} {2}},\ \bibinfo
  {pages} {658} (\bibinfo {year} {2012})},\ \Eprint
  {https://arxiv.org/abs/https://royalsocietypublishing.org/doi/pdf/10.1098/rsfs.2012.0023}
  {https://royalsocietypublishing.org/doi/pdf/10.1098/rsfs.2012.0023}
  \BibitemShut {NoStop}%
\bibitem [{Note1()}]{Note1}%
  \BibitemOpen
  \bibinfo {note} {There are $T$ inequivalent internal edges in a deltahedral
  tiling which are distributed into groupings of three (i.e. closed triangles).
  The minimal number of distinct triplets is $\lceil T/3 \rceil $.}\BibitemShut
  {Stop}%
\bibitem [{\citenamefont {Schroeder}\ \emph {et~al.}(2003)\citenamefont
  {Schroeder}, \citenamefont {Ramsden}, \citenamefont {Christy},\ and\
  \citenamefont {Hyde}}]{Turk2003}%
  \BibitemOpen
  \bibfield  {author} {\bibinfo {author} {\bibfnamefont {G.~E.}\ \bibnamefont
  {Schroeder}}, \bibinfo {author} {\bibfnamefont {S.~J.}\ \bibnamefont
  {Ramsden}}, \bibinfo {author} {\bibfnamefont {A.~G.}\ \bibnamefont
  {Christy}},\ and\ \bibinfo {author} {\bibfnamefont {S.~T.}\ \bibnamefont
  {Hyde}},\ }\bibfield  {title} {\bibinfo {title} {Medial surfaces of
  hyperbolic structures},\ }\href {https://doi.org/10.1140/epjb/e2003-00308-y}
  {\bibfield  {journal} {\bibinfo  {journal} {The European Physical Journal B -
  Condensed Matter}\ }\textbf {\bibinfo {volume} {35}},\ \bibinfo {pages} {551}
  (\bibinfo {year} {2003})}\BibitemShut {NoStop}%
\bibitem [{\citenamefont {Dotera}\ \emph {et~al.}(2017)\citenamefont {Dotera},
  \citenamefont {Tanaka},\ and\ \citenamefont {Takahashi}}]{Dotera2017}%
  \BibitemOpen
  \bibfield  {author} {\bibinfo {author} {\bibfnamefont {T.}~\bibnamefont
  {Dotera}}, \bibinfo {author} {\bibfnamefont {H.}~\bibnamefont {Tanaka}},\
  and\ \bibinfo {author} {\bibfnamefont {Y.}~\bibnamefont {Takahashi}},\
  }\bibfield  {title} {{\bibinfo {title} {Hexagulation
  numbers: the magic numbers of equal spheres on triply periodic minimal
  surfaces}},\ }\href {https://doi.org/10.1007/s11224-016-0833-7} {\bibfield
  {journal} {\bibinfo  {journal} {Structural Chemistry}\ }\textbf {\bibinfo
  {volume} {28}},\ \bibinfo {pages} {105} (\bibinfo {year} {2017})}\BibitemShut
  {NoStop}%
\bibitem [{\citenamefont {Sadoc}\ and\ \citenamefont
  {Charvolin}(1989)}]{Sadoc1989}%
  \BibitemOpen
  \bibfield  {author} {\bibinfo {author} {\bibfnamefont {J.-F.}\ \bibnamefont
  {Sadoc}}\ and\ \bibinfo {author} {\bibfnamefont {J.}~\bibnamefont
  {Charvolin}},\ }\bibfield  {title} {\bibinfo {title} {Infinite periodic
  minimal surfaces and their crystallography in the hyperbolic plane},\ }\href
  {https://doi.org/https://doi.org/10.1107/S0108767388008438} {\bibfield
  {journal} {\bibinfo  {journal} {Acta Crystallographica Section A}\ }\textbf
  {\bibinfo {volume} {45}},\ \bibinfo {pages} {10} (\bibinfo {year} {1989})},\
  \Eprint
  {https://arxiv.org/abs/https://onlinelibrary.wiley.com/doi/pdf/10.1107/S0108767388008438}
  {https://onlinelibrary.wiley.com/doi/pdf/10.1107/S0108767388008438}
  \BibitemShut {NoStop}%
\bibitem [{\citenamefont {Ramsden}\ \emph {et~al.}(2009)\citenamefont
  {Ramsden}, \citenamefont {Robins},\ and\ \citenamefont {Hyde}}]{Ramsden2009}%
  \BibitemOpen
  \bibfield  {author} {\bibinfo {author} {\bibfnamefont {S.~J.}\ \bibnamefont
  {Ramsden}}, \bibinfo {author} {\bibfnamefont {V.}~\bibnamefont {Robins}},\
  and\ \bibinfo {author} {\bibfnamefont {S.~T.}\ \bibnamefont {Hyde}},\
  }\bibfield  {title} {\bibinfo {title} {{Three-dimensional Euclidean nets from
  two-dimensional hyperbolic tilings: kaleidoscopic examples}},\ }\href
  {https://doi.org/10.1107/S0108767308040592} {\bibfield  {journal} {\bibinfo
  {journal} {Acta Crystallographica Section A}\ }\textbf {\bibinfo {volume}
  {65}},\ \bibinfo {pages} {81} (\bibinfo {year} {2009})}\BibitemShut {NoStop}%
\bibitem [{\citenamefont {Pedersen}\ and\ \citenamefont
  {Hyde}(2017)}]{Pedersen2017}%
  \BibitemOpen
  \bibfield  {author} {\bibinfo {author} {\bibfnamefont {M.~C.}\ \bibnamefont
  {Pedersen}}\ and\ \bibinfo {author} {\bibfnamefont {S.~T.}\ \bibnamefont
  {Hyde}},\ }\bibfield  {title} {\bibinfo {title} {{Hyperbolic crystallography
  of two-periodic surfaces and associated structures}},\ }\href
  {https://doi.org/10.1107/S2053273316019112} {\bibfield  {journal} {\bibinfo
  {journal} {Acta Crystallographica Section A}\ }\textbf {\bibinfo {volume}
  {73}},\ \bibinfo {pages} {124} (\bibinfo {year} {2017})}\BibitemShut
  {NoStop}%
\bibitem [{\citenamefont {Pedersen}\ \emph {et~al.}(2023)\citenamefont
  {Pedersen}, \citenamefont {Hyde}, \citenamefont {Ramsden},\ and\
  \citenamefont {Kirkensgaard}}]{Pedersen2023}%
  \BibitemOpen
  \bibfield  {author} {\bibinfo {author} {\bibfnamefont {M.~C.}\ \bibnamefont
  {Pedersen}}, \bibinfo {author} {\bibfnamefont {S.~T.}\ \bibnamefont {Hyde}},
  \bibinfo {author} {\bibfnamefont {S.}~\bibnamefont {Ramsden}},\ and\ \bibinfo
  {author} {\bibfnamefont {J.~J.~K.}\ \bibnamefont {Kirkensgaard}},\ }\bibfield
   {title} {\bibinfo {title} {Mapping hyperbolic order in curved materials},\
  }\href {https://doi.org/10.1039/D2SM01403C} {\bibfield  {journal} {\bibinfo
  {journal} {Soft Matter}\ }\textbf {\bibinfo {volume} {19}},\ \bibinfo {pages}
  {1586} (\bibinfo {year} {2023})}\BibitemShut {NoStop}%
\bibitem [{Note2()}]{Note2}%
  \BibitemOpen
  \bibinfo {note} {Note that, strictly, the triangulation in the hyperplane
  satisfies rotoinversion symmetries on the vertices ($\protect \bar {4}$) and
  centers ($\protect \bar {3}$) of the fundamental hexagonal cells as annotated
  Supporting Fig. S1A.}\BibitemShut {Stop}%
\bibitem [{\citenamefont {Chantler}\ \emph {et~al.}(2020)\citenamefont
  {Chantler}, \citenamefont {Boscherini},\ and\ \citenamefont
  {Bunker}}]{IntlTables}%
  \BibitemOpen
  \bibinfo {editor} {\bibfnamefont {C.~T.}\ \bibnamefont {Chantler}}, \bibinfo
  {editor} {\bibfnamefont {F.}~\bibnamefont {Boscherini}},\ and\ \bibinfo
  {editor} {\bibfnamefont {B.}~\bibnamefont {Bunker}},\ eds.,\ \href
  {https://doi.org/10.1107/97809553602060000116} {\emph {\bibinfo {title}
  {International {Tables} for {Crystallography}: {X}-ray absorption
  spectroscopy and related techniques}}},\ \bibinfo {edition} {1st}\ ed.,\
  Vol.~\bibinfo {volume} {I}\ (\bibinfo  {publisher} {International Union of
  Crystallography},\ \bibinfo {address} {Chester, England},\ \bibinfo {year}
  {2020})\BibitemShut {NoStop}%
\bibitem [{\citenamefont {Tyukodi}\ \emph {et~al.}(2022)\citenamefont
  {Tyukodi}, \citenamefont {Mohajerani}, \citenamefont {Hall}, \citenamefont
  {Grason},\ and\ \citenamefont {Hagan}}]{Tyukodi2022}%
  \BibitemOpen
  \bibfield  {author} {\bibinfo {author} {\bibfnamefont {B.}~\bibnamefont
  {Tyukodi}}, \bibinfo {author} {\bibfnamefont {F.}~\bibnamefont {Mohajerani}},
  \bibinfo {author} {\bibfnamefont {D.~M.}\ \bibnamefont {Hall}}, \bibinfo
  {author} {\bibfnamefont {G.~M.}\ \bibnamefont {Grason}},\ and\ \bibinfo
  {author} {\bibfnamefont {M.~F.}\ \bibnamefont {Hagan}},\ }\bibfield  {title}
  {\bibinfo {title} {{Thermodynamic Size Control in Curvature-Frustrated
  Tubules: Self-Limitation with Open Boundaries}},\ }\bibfield  {journal}
  {\bibinfo  {journal} {ACS Nano}\ }\href
  {https://doi.org/10.1021/acsnano.2c00865} {10.1021/acsnano.2c00865} (\bibinfo
  {year} {2022}),\ \Eprint {https://arxiv.org/abs/2109.01174}
  {arXiv:2109.01174} \BibitemShut {NoStop}%
\bibitem [{\citenamefont {Rotskoff}\ and\ \citenamefont
  {Geissler}(2018)}]{Rotskoff2018}%
  \BibitemOpen
  \bibfield  {author} {\bibinfo {author} {\bibfnamefont {G.~M.}\ \bibnamefont
  {Rotskoff}}\ and\ \bibinfo {author} {\bibfnamefont {P.~L.}\ \bibnamefont
  {Geissler}},\ }\bibfield  {title} {\bibinfo {title} {{Robust nonequilibrium
  pathways to microcompartment assembly}},\ }\href
  {https://doi.org/10.1073/pnas.1802499115} {\bibfield  {journal} {\bibinfo
  {journal} {Proceedings of the National Academy of Sciences of the United
  States of America}\ }\textbf {\bibinfo {volume} {115}},\ \bibinfo {pages}
  {6341} (\bibinfo {year} {2018})},\ \Eprint {https://arxiv.org/abs/1709.00321}
  {arXiv:1709.00321} \BibitemShut {NoStop}%
\bibitem [{\citenamefont {Li}\ \emph {et~al.}(2018)\citenamefont {Li},
  \citenamefont {Roy}, \citenamefont {Travesset},\ and\ \citenamefont
  {Zandi}}]{Li2018}%
  \BibitemOpen
  \bibfield  {author} {\bibinfo {author} {\bibfnamefont {S.}~\bibnamefont
  {Li}}, \bibinfo {author} {\bibfnamefont {P.}~\bibnamefont {Roy}}, \bibinfo
  {author} {\bibfnamefont {A.}~\bibnamefont {Travesset}},\ and\ \bibinfo
  {author} {\bibfnamefont {R.}~\bibnamefont {Zandi}},\ }\bibfield  {title}
  {\bibinfo {title} {Why large icosahedral viruses need scaffolding proteins},\
  }\href@noop {} {\bibfield  {journal} {\bibinfo  {journal} {Proc. Natl. Acad.
  Sci. U. S. A.}\ }\textbf {\bibinfo {volume} {115}},\ \bibinfo {pages} {10971}
  (\bibinfo {year} {2018})}\BibitemShut {NoStop}%
\bibitem [{\citenamefont {Panahandeh}\ \emph {et~al.}(2020)\citenamefont
  {Panahandeh}, \citenamefont {Li}, \citenamefont {Marichal}, \citenamefont
  {Leite~Rubim}, \citenamefont {Tresset},\ and\ \citenamefont
  {Zandi}}]{Panahandeh2020}%
  \BibitemOpen
  \bibfield  {author} {\bibinfo {author} {\bibfnamefont {S.}~\bibnamefont
  {Panahandeh}}, \bibinfo {author} {\bibfnamefont {S.}~\bibnamefont {Li}},
  \bibinfo {author} {\bibfnamefont {L.}~\bibnamefont {Marichal}}, \bibinfo
  {author} {\bibfnamefont {R.}~\bibnamefont {Leite~Rubim}}, \bibinfo {author}
  {\bibfnamefont {G.}~\bibnamefont {Tresset}},\ and\ \bibinfo {author}
  {\bibfnamefont {R.}~\bibnamefont {Zandi}},\ }\bibfield  {title} {\bibinfo
  {title} {How a virus circumvents energy barriers to form symmetric shells},\
  }\href {https://doi.org/10.1021/acsnano.9b08354} {\bibfield  {journal}
  {\bibinfo  {journal} {ACS Nano}\ }\textbf {\bibinfo {volume} {14}},\ \bibinfo
  {pages} {3170} (\bibinfo {year} {2020})}\BibitemShut {NoStop}%
\bibitem [{\citenamefont {Fang}\ \emph {et~al.}(2022)\citenamefont {Fang},
  \citenamefont {Tyukodi}, \citenamefont {Rogers},\ and\ \citenamefont
  {Hagan}}]{Fang2022}%
  \BibitemOpen
  \bibfield  {author} {\bibinfo {author} {\bibfnamefont {H.}~\bibnamefont
  {Fang}}, \bibinfo {author} {\bibfnamefont {B.}~\bibnamefont {Tyukodi}},
  \bibinfo {author} {\bibfnamefont {W.~B.}\ \bibnamefont {Rogers}},\ and\
  \bibinfo {author} {\bibfnamefont {M.~F.}\ \bibnamefont {Hagan}},\ }\bibfield
  {title} {\bibinfo {title} {{Polymorphic self-assembly of helical tubules is
  kinetically controlled}},\ }\href {https://doi.org/10.1039/d2sm00679k}
  {\bibfield  {journal} {\bibinfo  {journal} {Soft Matter}\ }\textbf {\bibinfo
  {volume} {18}},\ \bibinfo {pages} {6716} (\bibinfo {year} {2022})},\ \Eprint
  {https://arxiv.org/abs/2205.11598} {arXiv:2205.11598} \BibitemShut {NoStop}%
\bibitem [{\citenamefont {Mohajerani}\ \emph {et~al.}(2022)\citenamefont
  {Mohajerani}, \citenamefont {Tyukodi}, \citenamefont {Schlicksup},
  \citenamefont {Hadden-Perilla}, \citenamefont {Zlotnick},\ and\ \citenamefont
  {Hagan}}]{Mohajerani2022}%
  \BibitemOpen
  \bibfield  {author} {\bibinfo {author} {\bibfnamefont {F.}~\bibnamefont
  {Mohajerani}}, \bibinfo {author} {\bibfnamefont {B.}~\bibnamefont {Tyukodi}},
  \bibinfo {author} {\bibfnamefont {C.~J.}\ \bibnamefont {Schlicksup}},
  \bibinfo {author} {\bibfnamefont {J.~A.}\ \bibnamefont {Hadden-Perilla}},
  \bibinfo {author} {\bibfnamefont {A.}~\bibnamefont {Zlotnick}},\ and\
  \bibinfo {author} {\bibfnamefont {M.~F.}\ \bibnamefont {Hagan}},\ }\bibfield
  {title} {\bibinfo {title} {Multiscale modeling of hepatitis b virus capsid
  assembly and its dimorphism},\ }\href
  {https://doi.org/10.1021/acsnano.2c02119} {\bibfield  {journal} {\bibinfo
  {journal} {ACS Nano}\ }\textbf {\bibinfo {volume} {16}},\ \bibinfo {pages}
  {13845} (\bibinfo {year} {2022})},\ \bibinfo {note} {doi:
  10.1021/acsnano.2c02119}\BibitemShut {NoStop}%
\bibitem [{\citenamefont {Wagner}\ and\ \citenamefont
  {Zandi}(2015)}]{Wagner2015a}%
  \BibitemOpen
  \bibfield  {author} {\bibinfo {author} {\bibfnamefont {J.}~\bibnamefont
  {Wagner}}\ and\ \bibinfo {author} {\bibfnamefont {R.}~\bibnamefont {Zandi}},\
  }\bibfield  {title} {\bibinfo {title} {The robust assembly of small symmetric
  nanoshells},\ }\href@noop {} {\bibfield  {journal} {\bibinfo  {journal}
  {Biophys. J.}\ }\textbf {\bibinfo {volume} {109}},\ \bibinfo {pages} {956}
  (\bibinfo {year} {2015})}\BibitemShut {NoStop}%
\bibitem [{\citenamefont {Li}\ \emph {et~al.}(2019)\citenamefont {Li},
  \citenamefont {Zandi}, \citenamefont {Travesset},\ and\ \citenamefont
  {Grason}}]{Li2019}%
  \BibitemOpen
  \bibfield  {author} {\bibinfo {author} {\bibfnamefont {S.}~\bibnamefont
  {Li}}, \bibinfo {author} {\bibfnamefont {R.}~\bibnamefont {Zandi}}, \bibinfo
  {author} {\bibfnamefont {A.}~\bibnamefont {Travesset}},\ and\ \bibinfo
  {author} {\bibfnamefont {G.~M.}\ \bibnamefont {Grason}},\ }\bibfield  {title}
  {\bibinfo {title} {{Ground States of Crystalline Caps: Generalized Jellium on
  Curved Space}},\ }\bibfield  {journal} {\bibinfo  {journal} {Phys. Rev.
  Lett.}\ }\href {https://doi.org/10.1103/PhysRevLett.123.145501}
  {10.1103/PhysRevLett.123.145501} (\bibinfo {year} {2019}),\ \Eprint
  {https://arxiv.org/abs/1906.03301} {arXiv:1906.03301} \BibitemShut {NoStop}%
\bibitem [{\citenamefont {Schwartz}\ \emph {et~al.}(1998)\citenamefont
  {Schwartz}, \citenamefont {Shor}, \citenamefont {Prevelige},\ and\
  \citenamefont {Berger}}]{Schwartz1998}%
  \BibitemOpen
  \bibfield  {author} {\bibinfo {author} {\bibfnamefont {R.}~\bibnamefont
  {Schwartz}}, \bibinfo {author} {\bibfnamefont {P.~W.}\ \bibnamefont {Shor}},
  \bibinfo {author} {\bibfnamefont {P.~E.}\ \bibnamefont {Prevelige}},\ and\
  \bibinfo {author} {\bibfnamefont {B.}~\bibnamefont {Berger}},\ }\bibfield
  {title} {\bibinfo {title} {{Local Rules Simulation of the Kinetics of Virus
  Capsid Self-Assembly}},\ }\href@noop {} {\bibfield  {journal} {\bibinfo
  {journal} {Biophys. J.}\ }\textbf {\bibinfo {volume} {75}},\ \bibinfo {pages}
  {2626} (\bibinfo {year} {1998})}\BibitemShut {NoStop}%
\bibitem [{\citenamefont {Hagan}\ and\ \citenamefont
  {Chandler}(2006)}]{Hagan2006}%
  \BibitemOpen
  \bibfield  {author} {\bibinfo {author} {\bibfnamefont {M.~F.}\ \bibnamefont
  {Hagan}}\ and\ \bibinfo {author} {\bibfnamefont {D.}~\bibnamefont
  {Chandler}},\ }\bibfield  {title} {\bibinfo {title} {{Dynamic Pathways for
  Viral Capsid Assembly}},\ }\href@noop {} {\bibfield  {journal} {\bibinfo
  {journal} {Biophys. J.}\ }\textbf {\bibinfo {volume} {91}},\ \bibinfo {pages}
  {42} (\bibinfo {year} {2006})}\BibitemShut {NoStop}%
\bibitem [{\citenamefont {Jack}\ \emph {et~al.}(2007)\citenamefont {Jack},
  \citenamefont {Hagan},\ and\ \citenamefont {Chandler}}]{Jack2007}%
  \BibitemOpen
  \bibfield  {author} {\bibinfo {author} {\bibfnamefont {R.~L.}\ \bibnamefont
  {Jack}}, \bibinfo {author} {\bibfnamefont {M.~F.}\ \bibnamefont {Hagan}},\
  and\ \bibinfo {author} {\bibfnamefont {D.}~\bibnamefont {Chandler}},\
  }\bibfield  {title} {\bibinfo {title} {{Fluctuation-dissipation ratios in the
  dynamics of self-assembly}},\ }\href@noop {} {\bibfield  {journal} {\bibinfo
  {journal} {Phys. Rev. E}\ }\textbf {\bibinfo {volume} {76}},\ \bibinfo
  {pages} {021119} (\bibinfo {year} {2007})}\BibitemShut {NoStop}%
\bibitem [{\citenamefont {Hagan}\ \emph {et~al.}(2011)\citenamefont {Hagan},
  \citenamefont {Elrad},\ and\ \citenamefont {Jack}}]{Hagan2011}%
  \BibitemOpen
  \bibfield  {author} {\bibinfo {author} {\bibfnamefont {M.~F.}\ \bibnamefont
  {Hagan}}, \bibinfo {author} {\bibfnamefont {O.~M.}\ \bibnamefont {Elrad}},\
  and\ \bibinfo {author} {\bibfnamefont {R.~L.}\ \bibnamefont {Jack}},\
  }\bibfield  {title} {\bibinfo {title} {{Mechanisms of Kinetic Trapping in
  Self-Assembly and Phase Transformation}},\ }\href@noop {} {\bibfield
  {journal} {\bibinfo  {journal} {J. Chem. Phys.}\ }\textbf {\bibinfo {volume}
  {135}},\ \bibinfo {pages} {104115} (\bibinfo {year} {2011})}\BibitemShut
  {NoStop}%
\bibitem [{\citenamefont {Rapaport}(2008)}]{Rapaport2008}%
  \BibitemOpen
  \bibfield  {author} {\bibinfo {author} {\bibfnamefont {D.}~\bibnamefont
  {Rapaport}},\ }\bibfield  {title} {\bibinfo {title} {{The Role of
  Reversibility in Viral Capsid Growth: A Paradigm for Self-Assembly}},\
  }\href@noop {} {\bibfield  {journal} {\bibinfo  {journal} {Phys. Rev. Lett.}\
  }\textbf {\bibinfo {volume} {101}},\ \bibinfo {pages} {186101} (\bibinfo
  {year} {2008})}\BibitemShut {NoStop}%
\bibitem [{\citenamefont {Rapaport}(2010)}]{Rapaport2010}%
  \BibitemOpen
  \bibfield  {author} {\bibinfo {author} {\bibfnamefont {D.~C.}\ \bibnamefont
  {Rapaport}},\ }\bibfield  {title} {\bibinfo {title} {{Studies of reversible
  capsid shell growth}},\ }\href@noop {} {\bibfield  {journal} {\bibinfo
  {journal} {J. Phys.: Condens. Matter}\ }\textbf {\bibinfo {volume} {22}},\
  \bibinfo {pages} {104115} (\bibinfo {year} {2010})}\BibitemShut {NoStop}%
\bibitem [{\citenamefont {Rapaport}(2012)}]{Rapaport2012}%
  \BibitemOpen
  \bibfield  {author} {\bibinfo {author} {\bibfnamefont {D.~C.}\ \bibnamefont
  {Rapaport}},\ }\bibfield  {title} {\bibinfo {title} {{Molecular dynamics
  simulation of reversibly self-assembling shells in solution using trapezoidal
  particles}},\ }\href {https://doi.org/10.1103/PhysRevE.86.051917} {\bibfield
  {journal} {\bibinfo  {journal} {Phys. Rev. E}\ }\textbf {\bibinfo {volume}
  {86}},\ \bibinfo {pages} {051917} (\bibinfo {year} {2012})}\BibitemShut
  {NoStop}%
\bibitem [{\citenamefont {Whitelam}\ \emph {et~al.}(2009)\citenamefont
  {Whitelam}, \citenamefont {Rogers}, \citenamefont {Pasqua}, \citenamefont
  {Paavola}, \citenamefont {Trent},\ and\ \citenamefont
  {Geissler}}]{Whitelam2009}%
  \BibitemOpen
  \bibfield  {author} {\bibinfo {author} {\bibfnamefont {S.}~\bibnamefont
  {Whitelam}}, \bibinfo {author} {\bibfnamefont {C.}~\bibnamefont {Rogers}},
  \bibinfo {author} {\bibfnamefont {A.}~\bibnamefont {Pasqua}}, \bibinfo
  {author} {\bibfnamefont {C.}~\bibnamefont {Paavola}}, \bibinfo {author}
  {\bibfnamefont {J.}~\bibnamefont {Trent}},\ and\ \bibinfo {author}
  {\bibfnamefont {P.~L.}\ \bibnamefont {Geissler}},\ }\bibfield  {title}
  {\bibinfo {title} {{The Impact of Conformational Fluctuations on
  Self-Assembly: Cooperative Aggregation of Archaeal Chaperonin Proteins}},\
  }\href@noop {} {\bibfield  {journal} {\bibinfo  {journal} {Nano Lett.}\
  }\textbf {\bibinfo {volume} {9}},\ \bibinfo {pages} {292} (\bibinfo {year}
  {2009})}\BibitemShut {NoStop}%
\bibitem [{\citenamefont {Nguyen}\ \emph {et~al.}(2007)\citenamefont {Nguyen},
  \citenamefont {Reddy},\ and\ \citenamefont {Brooks}}]{Nguyen2007}%
  \BibitemOpen
  \bibfield  {author} {\bibinfo {author} {\bibfnamefont {H.~D.}\ \bibnamefont
  {Nguyen}}, \bibinfo {author} {\bibfnamefont {V.~S.}\ \bibnamefont {Reddy}},\
  and\ \bibinfo {author} {\bibfnamefont {C.~L.}\ \bibnamefont {Brooks}},\
  }\bibfield  {title} {\bibinfo {title} {{Deciphering the Kinetic Mechanism of
  Spontaneous Self-Assembly of Icosahedral Capsids}},\ }\href@noop {}
  {\bibfield  {journal} {\bibinfo  {journal} {Nano Lett.}\ }\textbf {\bibinfo
  {volume} {7}},\ \bibinfo {pages} {338} (\bibinfo {year} {2007})}\BibitemShut
  {NoStop}%
\bibitem [{\citenamefont {Nguyen}\ and\ \citenamefont
  {Brooks}(2008)}]{Nguyen2008}%
  \BibitemOpen
  \bibfield  {author} {\bibinfo {author} {\bibfnamefont {H.}~\bibnamefont
  {Nguyen}}\ and\ \bibinfo {author} {\bibfnamefont {C.}~\bibnamefont
  {Brooks}},\ }\bibfield  {title} {\bibinfo {title} {{Generalized structural
  polymorphism in self-assembled viral particles}},\ }\href@noop {} {\bibfield
  {journal} {\bibinfo  {journal} {Nano Lett.}\ }\textbf {\bibinfo {volume}
  {8}},\ \bibinfo {pages} {4574} (\bibinfo {year} {2008})}\BibitemShut
  {NoStop}%
\bibitem [{\citenamefont {Nguyen}\ \emph {et~al.}(2009)\citenamefont {Nguyen},
  \citenamefont {Reddy},\ and\ \citenamefont {Brooks}}]{Nguyen2009}%
  \BibitemOpen
  \bibfield  {author} {\bibinfo {author} {\bibfnamefont {H.~D.}\ \bibnamefont
  {Nguyen}}, \bibinfo {author} {\bibfnamefont {V.~S.}\ \bibnamefont {Reddy}},\
  and\ \bibinfo {author} {\bibfnamefont {C.~L.}\ \bibnamefont {Brooks}},\
  }\bibfield  {title} {\bibinfo {title} {{Invariant Polymorphism in Virus
  Capsid Assembly}},\ }\href@noop {} {\bibfield  {journal} {\bibinfo  {journal}
  {J. Am. Chem. Soc.}\ }\textbf {\bibinfo {volume} {131}},\ \bibinfo {pages}
  {2606} (\bibinfo {year} {2009})}\BibitemShut {NoStop}%
\bibitem [{\citenamefont {Wilber}\ \emph {et~al.}(2007)\citenamefont {Wilber},
  \citenamefont {Doye}, \citenamefont {Louis}, \citenamefont {Noya},
  \citenamefont {Miller},\ and\ \citenamefont {Wong}}]{Wilber2007}%
  \BibitemOpen
  \bibfield  {author} {\bibinfo {author} {\bibfnamefont {A.~W.}\ \bibnamefont
  {Wilber}}, \bibinfo {author} {\bibfnamefont {J.~P.~K.}\ \bibnamefont {Doye}},
  \bibinfo {author} {\bibfnamefont {A.~A.}\ \bibnamefont {Louis}}, \bibinfo
  {author} {\bibfnamefont {E.~G.}\ \bibnamefont {Noya}}, \bibinfo {author}
  {\bibfnamefont {M.~A.}\ \bibnamefont {Miller}},\ and\ \bibinfo {author}
  {\bibfnamefont {P.}~\bibnamefont {Wong}},\ }\bibfield  {title} {\bibinfo
  {title} {{Reversible Self-Assembly of Patchy Particles into Monodisperse
  Icosahedral Clusters}},\ }\href@noop {} {\bibfield  {journal} {\bibinfo
  {journal} {J. Chem. Phys.}\ }\textbf {\bibinfo {volume} {127}},\ \bibinfo
  {pages} {085106} (\bibinfo {year} {2007})}\BibitemShut {NoStop}%
\bibitem [{\citenamefont {Wilber}\ \emph {et~al.}(2009)\citenamefont {Wilber},
  \citenamefont {Doye}, \citenamefont {Louis},\ and\ \citenamefont
  {Lewis}}]{Wilber2009}%
  \BibitemOpen
  \bibfield  {author} {\bibinfo {author} {\bibfnamefont {A.~W.}\ \bibnamefont
  {Wilber}}, \bibinfo {author} {\bibfnamefont {J.~P.~K.}\ \bibnamefont {Doye}},
  \bibinfo {author} {\bibfnamefont {A.~A.}\ \bibnamefont {Louis}},\ and\
  \bibinfo {author} {\bibfnamefont {A.~C.~F.}\ \bibnamefont {Lewis}},\
  }\bibfield  {title} {\bibinfo {title} {{Monodisperse self-assembly in a model
  with protein-like interactions}},\ }\href@noop {} {\bibfield  {journal}
  {\bibinfo  {journal} {J. Chem. Phys.}\ }\textbf {\bibinfo {volume} {131}},\
  \bibinfo {pages} {175102} (\bibinfo {year} {2009})}\BibitemShut {NoStop}%
\bibitem [{\citenamefont {Cheng}\ \emph {et~al.}(2012)\citenamefont {Cheng},
  \citenamefont {Aggarwal},\ and\ \citenamefont {Stevens}}]{Cheng2012}%
  \BibitemOpen
  \bibfield  {author} {\bibinfo {author} {\bibfnamefont {S.}~\bibnamefont
  {Cheng}}, \bibinfo {author} {\bibfnamefont {A.}~\bibnamefont {Aggarwal}},\
  and\ \bibinfo {author} {\bibfnamefont {M.~J.}\ \bibnamefont {Stevens}},\
  }\bibfield  {title} {\bibinfo {title} {{Self-assembly of artificial
  microtubules}},\ }\href {https://doi.org/10.1039/c2sm25068c} {\bibfield
  {journal} {\bibinfo  {journal} {Soft Matter}\ }\textbf {\bibinfo {volume}
  {8}},\ \bibinfo {pages} {5666} (\bibinfo {year} {2012})},\ \Eprint
  {https://arxiv.org/abs/arXiv:1201.2328v1} {arXiv:arXiv:1201.2328v1}
  \BibitemShut {NoStop}%
\bibitem [{\citenamefont {Hagan}(2014)}]{Hagan2014}%
  \BibitemOpen
  \bibfield  {author} {\bibinfo {author} {\bibfnamefont {M.~F.}\ \bibnamefont
  {Hagan}},\ }\bibfield  {title} {\bibinfo {title} {{Modeling Viral Capsid
  Assembly}},\ }\href@noop {} {\bibfield  {journal} {\bibinfo  {journal} {Adv.
  Chem. Phys.}\ }\textbf {\bibinfo {volume} {155}},\ \bibinfo {pages} {1}
  (\bibinfo {year} {2014})}\BibitemShut {NoStop}%
\bibitem [{\citenamefont {Whitelam}\ and\ \citenamefont
  {Jack}(2015)}]{Whitelam2015}%
  \BibitemOpen
  \bibfield  {author} {\bibinfo {author} {\bibfnamefont {S.}~\bibnamefont
  {Whitelam}}\ and\ \bibinfo {author} {\bibfnamefont {R.~L.}\ \bibnamefont
  {Jack}},\ }\bibfield  {title} {\bibinfo {title} {{The Statistical Mechanics
  of Dynamic Pathways to Self-assembly}},\ }\href@noop {} {\bibfield  {journal}
  {\bibinfo  {journal} {Ann Rev Phys Chem}\ }\textbf {\bibinfo {volume} {66}},\
  \bibinfo {pages} {143} (\bibinfo {year} {2015})}\BibitemShut {NoStop}%
\bibitem [{\citenamefont {Seung}\ and\ \citenamefont
  {Nelson}(1988)}]{Seung1988}%
  \BibitemOpen
  \bibfield  {author} {\bibinfo {author} {\bibfnamefont {H.~S.}\ \bibnamefont
  {Seung}}\ and\ \bibinfo {author} {\bibfnamefont {D.~R.}\ \bibnamefont
  {Nelson}},\ }\bibfield  {title} {\bibinfo {title} {Defects in flexible
  membranes with crystalline order},\ }\href
  {https://doi.org/10.1103/PhysRevA.38.1005} {\bibfield  {journal} {\bibinfo
  {journal} {Phys. Rev. A}\ }\textbf {\bibinfo {volume} {38}},\ \bibinfo
  {pages} {1005} (\bibinfo {year} {1988})}\BibitemShut {NoStop}%
\bibitem [{Note3()}]{Note3}%
  \BibitemOpen
  \bibinfo {note} {Our definition of disclination charge accounts for the
  effective negative Gaussian curvature of the target crystal, but defines
  excess bond rotation relative to a target bond coordination that may be
  larger than 6 (e.g. the 8-coordinated vertex at position 24d of G
  structures).}\BibitemShut {Stop}%
\bibitem [{\citenamefont {Grason}(2012)}]{Grason2012}%
  \BibitemOpen
  \bibfield  {author} {\bibinfo {author} {\bibfnamefont {G.~M.}\ \bibnamefont
  {Grason}},\ }\bibfield  {title} {\bibinfo {title} {Defects in crystalline
  packings of twisted filament bundles. i. continuum theory of disclinations},\
  }\href {https://doi.org/10.1103/PhysRevE.85.031603} {\bibfield  {journal}
  {\bibinfo  {journal} {Phys. Rev. E}\ }\textbf {\bibinfo {volume} {85}},\
  \bibinfo {pages} {031603} (\bibinfo {year} {2012})}\BibitemShut {NoStop}%
\bibitem [{\citenamefont {Schneider}\ and\ \citenamefont
  {Gompper}(2005)}]{Schneider2005}%
  \BibitemOpen
  \bibfield  {author} {\bibinfo {author} {\bibfnamefont {S.}~\bibnamefont
  {Schneider}}\ and\ \bibinfo {author} {\bibfnamefont {G.}~\bibnamefont
  {Gompper}},\ }\bibfield  {title} {\bibinfo {title} {Shapes of crystalline
  domains on spherical fluid vesicles},\ }\href
  {https://doi.org/10.1209/epl/i2004-10464-2} {\bibfield  {journal} {\bibinfo
  {journal} {Europhysics Letters}\ }\textbf {\bibinfo {volume} {70}},\ \bibinfo
  {pages} {136} (\bibinfo {year} {2005})}\BibitemShut {NoStop}%
\bibitem [{\citenamefont {Meng}\ \emph {et~al.}(2014)\citenamefont {Meng},
  \citenamefont {Paulose}, \citenamefont {Nelson},\ and\ \citenamefont
  {Manoharan}}]{Meng2014}%
  \BibitemOpen
  \bibfield  {author} {\bibinfo {author} {\bibfnamefont {G.}~\bibnamefont
  {Meng}}, \bibinfo {author} {\bibfnamefont {J.}~\bibnamefont {Paulose}},
  \bibinfo {author} {\bibfnamefont {D.~R.}\ \bibnamefont {Nelson}},\ and\
  \bibinfo {author} {\bibfnamefont {V.~N.}\ \bibnamefont {Manoharan}},\
  }\bibfield  {title} {\bibinfo {title} {{E}lastic {I}nstability of a {C}rystal
  {G}rowing on a {C}urved {S}urface},\ }\href@noop {} {\bibfield  {journal}
  {\bibinfo  {journal} {Science}\ }\textbf {\bibinfo {volume} {343}} (\bibinfo
  {year} {2014})}\BibitemShut {NoStop}%
\bibitem [{\citenamefont {Grason}(2016)}]{Grason2016}%
  \BibitemOpen
  \bibfield  {author} {\bibinfo {author} {\bibfnamefont {G.~M.}\ \bibnamefont
  {Grason}},\ }\bibfield  {title} {\bibinfo {title} {Perspective: Geometrically
  frustrated assemblies},\ }\href {https://doi.org/10.1063/1.4962629}
  {\bibfield  {journal} {\bibinfo  {journal} {J. Chem. Phys.}\ }\textbf
  {\bibinfo {volume} {145}},\ \bibinfo {pages} {110901} (\bibinfo {year}
  {2016})},\ \Eprint
  {https://arxiv.org/abs/http://dx.doi.org/10.1063/1.4962629}
  {http://dx.doi.org/10.1063/1.4962629} \BibitemShut {NoStop}%
\bibitem [{\citenamefont {Yang}\ \emph {et~al.}(2010)\citenamefont {Yang},
  \citenamefont {Meyer},\ and\ \citenamefont {Hagan}}]{Yang2010}%
  \BibitemOpen
  \bibfield  {author} {\bibinfo {author} {\bibfnamefont {Y.}~\bibnamefont
  {Yang}}, \bibinfo {author} {\bibfnamefont {R.}~\bibnamefont {Meyer}},\ and\
  \bibinfo {author} {\bibfnamefont {M.~F.}\ \bibnamefont {Hagan}},\ }\bibfield
  {title} {\bibinfo {title} {{Self-Limited Self-Assembly of Chiral
  Filaments}},\ }\href@noop {} {\bibfield  {journal} {\bibinfo  {journal}
  {Phys. Rev. Lett.}\ }\textbf {\bibinfo {volume} {104}},\ \bibinfo {pages}
  {258102} (\bibinfo {year} {2010})}\BibitemShut {NoStop}%
\bibitem [{\citenamefont {Saranathan}\ \emph
  {et~al.}(2021{\natexlab{b}})\citenamefont {Saranathan}, \citenamefont
  {Narayanan}, \citenamefont {Sandy}, \citenamefont {Dufresne},\ and\
  \citenamefont {Prum}}]{Saranathan2011}%
  \BibitemOpen
  \bibfield  {author} {\bibinfo {author} {\bibfnamefont {V.}~\bibnamefont
  {Saranathan}}, \bibinfo {author} {\bibfnamefont {S.}~\bibnamefont
  {Narayanan}}, \bibinfo {author} {\bibfnamefont {A.}~\bibnamefont {Sandy}},
  \bibinfo {author} {\bibfnamefont {E.~R.}\ \bibnamefont {Dufresne}},\ and\
  \bibinfo {author} {\bibfnamefont {R.~O.}\ \bibnamefont {Prum}},\ }\bibfield
  {title} {\bibinfo {title} {Evolution of single gyroid photonic crystals in
  bird feathers},\ }\href {https://doi.org/10.1073/pnas.2101357118} {\bibfield
  {journal} {\bibinfo  {journal} {Proceedings of the National Academy of
  Sciences}\ }\textbf {\bibinfo {volume} {118}},\ \bibinfo {pages}
  {e2101357118} (\bibinfo {year} {2021}{\natexlab{b}})},\ \Eprint
  {https://arxiv.org/abs/https://www.pnas.org/doi/pdf/10.1073/pnas.2101357118}
  {https://www.pnas.org/doi/pdf/10.1073/pnas.2101357118} \BibitemShut {NoStop}%
\bibitem [{\citenamefont {Dowling}\ \emph {et~al.}(2023)\citenamefont
  {Dowling}, \citenamefont {Park}, \citenamefont {Gerstenmaier}, \citenamefont
  {Yang}, \citenamefont {Wargacki}, \citenamefont {Hsia}, \citenamefont
  {Fries}, \citenamefont {Ravichandran}, \citenamefont {Walkey}, \citenamefont
  {Burrell}, \citenamefont {Veesler}, \citenamefont {Baker},\ and\
  \citenamefont {King}}]{Dowling2023}%
  \BibitemOpen
  \bibfield  {author} {\bibinfo {author} {\bibfnamefont {Q.~M.}\ \bibnamefont
  {Dowling}}, \bibinfo {author} {\bibfnamefont {Y.-J.}\ \bibnamefont {Park}},
  \bibinfo {author} {\bibfnamefont {N.}~\bibnamefont {Gerstenmaier}}, \bibinfo
  {author} {\bibfnamefont {E.~C.}\ \bibnamefont {Yang}}, \bibinfo {author}
  {\bibfnamefont {A.}~\bibnamefont {Wargacki}}, \bibinfo {author}
  {\bibfnamefont {Y.}~\bibnamefont {Hsia}}, \bibinfo {author} {\bibfnamefont
  {C.~N.}\ \bibnamefont {Fries}}, \bibinfo {author} {\bibfnamefont
  {R.}~\bibnamefont {Ravichandran}}, \bibinfo {author} {\bibfnamefont
  {C.}~\bibnamefont {Walkey}}, \bibinfo {author} {\bibfnamefont
  {A.}~\bibnamefont {Burrell}}, \bibinfo {author} {\bibfnamefont
  {D.}~\bibnamefont {Veesler}}, \bibinfo {author} {\bibfnamefont
  {D.}~\bibnamefont {Baker}},\ and\ \bibinfo {author} {\bibfnamefont {N.~P.}\
  \bibnamefont {King}},\ }\href {https://doi.org/10.1101/2023.06.16.545393}
  {{\emph {\bibinfo {title} {Hierarchical design of
  pseudosymmetric protein nanoparticles}}}},\ \bibinfo {type} {preprint}\
  (\bibinfo  {institution} {Bioengineering},\ \bibinfo {year}
  {2023})\BibitemShut {NoStop}%
\bibitem [{\citenamefont {Zandi}\ \emph {et~al.}(2006)\citenamefont {Zandi},
  \citenamefont {van~der Schoot}, \citenamefont {Reguera}, \citenamefont
  {Kegel},\ and\ \citenamefont {Reiss}}]{Zandi2006}%
  \BibitemOpen
  \bibfield  {author} {\bibinfo {author} {\bibfnamefont {R.}~\bibnamefont
  {Zandi}}, \bibinfo {author} {\bibfnamefont {P.}~\bibnamefont {van~der
  Schoot}}, \bibinfo {author} {\bibfnamefont {D.}~\bibnamefont {Reguera}},
  \bibinfo {author} {\bibfnamefont {W.}~\bibnamefont {Kegel}},\ and\ \bibinfo
  {author} {\bibfnamefont {H.}~\bibnamefont {Reiss}},\ }\bibfield  {title}
  {\bibinfo {title} {{Classical Nucleation Theory of Virus Capsids}},\
  }\href@noop {} {\bibfield  {journal} {\bibinfo  {journal} {Biophys. J.}\
  }\textbf {\bibinfo {volume} {90}},\ \bibinfo {pages} {1939} (\bibinfo {year}
  {2006})}\BibitemShut {NoStop}%
\bibitem [{\citenamefont {Zandi}\ \emph {et~al.}(2020)\citenamefont {Zandi},
  \citenamefont {Dragnea}, \citenamefont {Travesset},\ and\ \citenamefont
  {Podgornik}}]{Zandi2020}%
  \BibitemOpen
  \bibfield  {author} {\bibinfo {author} {\bibfnamefont {R.}~\bibnamefont
  {Zandi}}, \bibinfo {author} {\bibfnamefont {B.}~\bibnamefont {Dragnea}},
  \bibinfo {author} {\bibfnamefont {A.}~\bibnamefont {Travesset}},\ and\
  \bibinfo {author} {\bibfnamefont {R.}~\bibnamefont {Podgornik}},\ }\bibfield
  {title} {\bibinfo {title} {On virus growth and form},\ }\href@noop {}
  {\bibfield  {journal} {\bibinfo  {journal} {Phys. Rep.}\ } (\bibinfo {year}
  {2020})}\BibitemShut {NoStop}%
\bibitem [{\citenamefont {sarah-marie belcastro}\ and\ \citenamefont
  {Hull}(2002)}]{Belcastro2002}%
  \BibitemOpen
  \bibfield  {author} {\bibinfo {author} {\bibnamefont {sarah-marie
  belcastro}}\ and\ \bibinfo {author} {\bibfnamefont {T.~C.}\ \bibnamefont
  {Hull}},\ }\bibfield  {title} {\bibinfo {title} {Modelling the folding of
  paper into three dimensions using affine transformations},\ }\href
  {https://doi.org/https://doi.org/10.1016/S0024-3795(01)00608-5} {\bibfield
  {journal} {\bibinfo  {journal} {Linear Algebra and its Applications}\
  }\textbf {\bibinfo {volume} {348}},\ \bibinfo {pages} {273} (\bibinfo {year}
  {2002})}\BibitemShut {NoStop}%
\end{thebibliography}%

\end{document}